\documentclass{ar-1col-S2O}

\usepackage[comma]{natbib}
\usepackage{url}
\usepackage[colorlinks=true,citecolor=cyan]{hyperref}
\usepackage{wrapfig}

\jname{Annual Reviews of Astronomy and Astrophysics}
\jvol{AA}
\jyear{2025}
\doi{10.1146/((please add article doi))}

\begin{document}
\hypersetup{linktoc=all,linkcolor=black} % To get Table of Contents color correct

% Page header
\markboth{Beuther, Kuiper \& Tafalla}{Star formation: Comparing low- to high-mass}

% Title
\title{Star formation from low to high mass:\\ A comparative view}

%Authors, affiliations address.
\author{H.~Beuther,$^1$ R.~Kuiper,$^2$ and M.~Tafalla$^3$
\affil{$^1$Max Planck Institute for Astronomy, K\"onigstuhl 17,
     69117 Heidelberg, Germany; email: beuther@mpia.de}
\affil{$^2$Faculty of Physics, University of Duisburg-Essen, Lotharstraße 1, 47057 Duisburg, Germany}
\affil{$^3$Observatorio Astronómico Nacional (IGN), Alfonso XII 3, 28014 Madrid, Spain}}

%Abstract
\begin{abstract}
\begin{minipage}[l]{0.75\textwidth}
Star formation has often been studied by separating the low- and high-mass regimes with an approximate boundary at 8\,$M_{\odot}$. While some of the outcomes of the star-formation process are different between the two regimes, it is less clear whether the physical processes leading to these outcomes are that different at all. Here, we systematically compare low- and high-mass star formation by reviewing the most important processes and quantities from an observational and theoretical point of view. We identify three regimes where processes are either similar, quantitatively or qualitatively different between low- and high-mass star formation.  
\begin{itemize}
\item Similar characteristics can be identified for the turbulent gas properties and density structures of the star-forming regions. Many of the observational characteristics also do not depend that strongly on the environment.
\item Quantitative differences can be found for outflow, infall and accretion rates as well as mean column and volume densities. Also the multiplicity significantly rises from low- to high-mass stars. The importance of the magnetic field for the formation processes appears still less well constrained.
\item Qualitative differences between low- and high-mass star formation relate mainly to the radiative and ionizing feedback that occurs almost exclusively in regions forming high-mass stars. Nevertheless, accretion apparently can continue via disk structures in ionized accretion flows.
\end{itemize}
Finally, we discuss to what extent a unified picture of star formation over all masses is possible and which issues need to be addressed in the future.

%Abstract text, approximately 225 words and 
%inclusive of 3--5 bullet items describing 
%the findings of the current research:
%\begin{itemize}
%\item First finding.
%\item Second finding.
%\item Third finding.
%\end{itemize}

\end{minipage}
\end{abstract}

%Keywords, etc.
\maketitle
%Table of Contents
\tableofcontents

% Heading 1
\section{Introduction}
\label{intro}

\subsection{Motivation}

Star formation has been a very active field of research since the end of the 1960s. Tremendous progress has been made since then in the understanding of the physical and chemical processes during the formation of stars over the entire mass range from $\sim$0.08\,M$_{\odot}$ (the lower stellar mass end at the brown dwarf boundary) to the most massive stars even exceeding 100\,M$_{\odot}$ (e.g., \citealt{crowther2010}). Several reviews over the past decades gave excellent summaries about the state-of-art at the given time, just to name a few: \citet{shu1987}, \citet{evans1999}, \citet{mckee2007}, \citet{arce2007}, \citet{bergin2007}, \citet{zinnecker2007}, \citet{beuther2007}, \citet{herbst2009}, \citet{andre2014}, \citet{offner2014}, \citet{motte2018}, \citet{megeath2022}, \citet{hacar2023}, \citet{pineda2023}, and \citet{tobin2024}. Furthermore, the Protostars and Planets conference and book series gave excellent reviews about diverse sub-aspects in the field (e.g., \citealt{reipurth2007}, \citealt{beuther2014}, \citealt{frank2014}, \citealt{inutsuka2023}).

A common practice in star-formation research has been to distinguish the formation of low- and high-mass stars, with the dividing line being approximately 8\,$M_{\odot}$. This separation is based on several ideas. From a star-formation perspective, (proto)stars larger than 8\,$M_{\odot}$ barely have an observable pre-main-sequence evolution in the Hertzsprung-Russell diagram but reach the zero-age-main-sequence while still accreting gas from the environmental envelope (e.g., \citealt{palla1993}). From a perspective of stellar physics, roughly an 8\,$M_{\odot}$ star is also needed to have a final central core mass of $\sim$1.4\,$M_{\odot}$ \citep[Chandrasekhar limit, ][]{1931ApJ....74...81C} that can lead to the formation of a neutron star or black hole. In addition to this, above 8\,$M_{\odot}$ the photon fluxes short-ward of the Lyman limit strongly increase which results in the formation of H{\sc ii} regions (e.g., \citealt{spitzer1998}). Such an 8\,$M_{\odot}$ main sequence star corresponds roughly to a B3 stellar type with a luminosity of $\sim$2$\times 10^3$\,L$_{\odot}$ \citep{lang1992}.

Because of these differences, low- and high-mass star formation present several basic phenomenological differences, e.g.: high-mass (proto)stars must still be accreting on the main-sequence although already hydrogen burning,  they form hypercompact H{\sc ii} regions (HCH{\sc ii}s), which accretion must penetrate (e.g., \citealt{keto2002,tan2003,beuther2007}), they form almost exclusively in clusters (e.g., \citealt{lada2003,zinnecker2007,tan2014,motte2018,megeath2022}), and they have a higher degree of multiplicity compared to their low-mass counterparts (e.g., \citealt{preibisch1999,lada2006,zinnecker2007,motte2018,offner2023}). 
In contrast to these differences, there are also many processes that should be qualitatively independent of the final stellar mass or only scale with the mass of the final star, e.g.: 
star formation is driven by gravity and modulated by magnetic fields and turbulence (e.g., \citealt{mckee2007,pattle2023}), 
the earliest phases of collapse and protostellar formation is marked by first and second hydrostatic Larson core formation due to phase transitions \citep{larson1969, zotero-2933, zotero-3450, zotero-2558, zotero-2909, 2023A&A...680A..23A}, qualitatively similar for low- and high-mass stars, but with a clear mass-scaling \citep{zotero-2558},
disk formation takes place over the entire mass regime (e.g., \citealt{dutrey2014,beltran2016,andrews2018,ahmadi2023}), 
molecular jets and entrained outflows are observed around protostars of all masses (e.g., \citealt{bontemps1996,beuther2002b,wu2004,beuther2005,arce2007,lopez2009,frank2014,maud2015}), and 
stars give rise to a continuous and almost universal initial mass function (IMF, e.g., \citealt{salpeter1955,corbelli2005,offner2014}). 
%In the following, we refer to low-mass regions as star-forming regions that form only low-mass stars (either isolated objects, or more typically distributed populations like Taurus), whereas high-mass regions are those that form clusters containing high-, intermediate- and low-mass protostars typically following an IMF distribution. 

\begin{marginnote}[]
\entry{Low-mass star formation}{Refers to either isolated low-mass cores or distributed star-forming regions with many/several protostars of low-mass only.}
\entry{High-mass star formation}{Refers always to clustered star formation that forms high- and low-mass stars together, typically following an initial mass function.}
\end{marginnote}

While most previous reviews of star formation have emphasized the dichotomy between low- and high-mass stars, this review aims to assess critically the similarities and differences between the two regimes, and to investigate the path toward a more unified description of star formation. To that end, we provide an overview of the current progress in our understanding of low- and high-mass star formation, and discuss the challenges remaining to bring together the work of the research communities that study each regime.

This review focuses on the formation of stars in typical Galactic regions (local environment and spiral arm/interarm regions), and excludes extreme environments like the Central Molecular Zone (CMZ) of the Milky Way or galactic mergers. It also restricts star formation to that resulting from the collapse of molecular gas, excluding second-generation processes such as mass transfer or stellar mergers, which represent the evolution of objects already formed via molecular-gas collapse.

%\begin{marginnote}[]
%\entry{Intermediate-mass stars}{Since we mainly compare low- with high-mass star formation with a separation around 8\,$M_{\odot}$ (section \ref{intro}), typical intermediate-mass Herbig Ae/Be stars of a few $M_{\odot}$ are usually considered here as part of the low-mass regime.}
%\end{marginnote}
\begin{textbox}[ht]\subsection{Clustering}
While many of the results for low-mass regions are based on isolated low-mass cores (e.g., \citealt{bergin2007}), the clustered properties of low-mass stars are also important (e.g., \citealt{lada2003,megeath2022}). In contrast, high-mass stars always form in clusters and hence multiple objects need to be taken into account (e.g., \citealt{beuther2007,zinnecker2007,tan2014, 2020SSRv..216...64K}). Although one often compares low-mass star formation in nearby regions like Taurus or Perseus with high-mass star formation, the majority of low-mass stars form simultaneously with the high-mass stars in the clustered environment of high-mass star-forming regions.
\subsection{Intermediate-mass stars}
Since we mainly compare low- with high-mass star formation with a separation around 8\,$M_{\odot}$ (section \ref{intro}), typical intermediate-mass Herbig Ae/Be stars of a few $M_{\odot}$ are usually considered here as part of the low-mass regime. These intermediate-mass stars are typically also found already in little clusters (e.g., \citealt{testi1999, 2023MNRAS.519.3958I}).
\end{textbox}

\subsection{Conceptual and historical differences between the two fields}

\begin{marginnote}[]
%\entry{Cores}{We talk about cores as condensations typically forming individual stars or small multiples like binaries, triplets, or quadruple systems. The core sizes depend on the densities and can be as large as $\sim$0.25\,pc for low-mass globules like B68 (e.g., \citealt{alves2001}), and smaller than 0.1\,pc down to a few thousand au for high-mass cores (e.g., for a comparison see \citealt{beuther2013}).}
\entry{Cores}{Condensations forming individual stars or small multiple systems. Sizes depend on densities and vary between $\sim$0.25\,pc for low-mass globules and $<$0.1\,pc down to a few 1000\,au for high-mass cores.}
%\entry{Clumps}{The term clump is used for the larger-scale cluster-forming condensations typically found in high-mass star formation on scales of $\sim$1\,pc (e.g., \citealt{beuther2007}). Such clumps can potentially form at hub-filament intersections (e.g., \citealt{kumar2020}).}
\entry{Clumps}{Larger-scale cluster-forming condensations typically found in high-mass star formation on scales of $\sim$1\,pc.}
%Such clumps can potentially form at hub-filament intersections (e.g., \citealt{kumar2020}).}
\end{marginnote}

The standard model of (low-mass) star formation has been developed since the seventies (e.g., \citealt{larson1969,shu1977,shu1987}) and is covered in excellent textbooks and reviews like \citet{stahler2005} or \citet{mckee2007}. The additional importance of turbulence was discussed intensively since the beginning of the new millenium (e.g., \citealt{klessen2000,maclow2004,federrath2015}). Here, we concentrate on conceptual and observational differences between low- and high-mass star formation as discussed over the last few decades. Because high-mass protostars reach the main sequence before they have finished accretion, strong radiation will affect the ongoing mass growth. This leaves one with the question of whether high-mass star formation is simply a scaled-up version of low-mass star formation, or whether this higher radiation leads to more fundamental differences in the formation process (e.g., \citealt{zinnecker2007}).

In spherical symmetry, \citet{kahn1974} argued that the radiation pressure onto dust grains limits the formation of stars more massive than 40\,M$_{\odot}$. \citet{wolfire1987} investigated the dust properties needed to allow accretion flows also for more massive stars, and they found that a significant alteration of standard dust properties \citep{mathis1977} is required to overcome the radiation pressure. However, such strong changes of the dust properties have so far not been observed in typical high-mass star-forming regions. 

To overcome the radiation limit, Bonnell and co-workers introduced a scenario
in a series of papers where competitive accretion within the cluster-forming environment plays a major role (e.g., \citealt{bonnell1998,bonnell2004,bonnell2007}). In this picture, the gas potential of the forming cluster is important, and protostars that reside in the deepest well of the gravitational potential can accrete most of the gas. In extremely dense cluster-forming potentials, protostellar mergers may occur (e.g., \citealt{bonnell2002,bonnell2005}). While the entire gas potential has certainly to contribute to the gas flow and by that also to the accretion processes, the general importance of competitive accretion has been investigated and debated at length (e.g., \citealt{bonnell2007,tan2014,vazquez2019,padoan2020}). Protostellar mergers may occur in exceptional cases, but they do not seem to be a major path for high-mass star formation (e.g., \citealt{zinnecker2007,bally2020}).

As an alternative scenario, \citet{mckee2002,mckee2003} developed the so-called turbulent core model (see also \citealt{tan2014}). While low-mass accretion rates are related to the thermal sound speed, \citet{mckee2003} argue that high-mass stars are forming in massive, turbulent supersonic cores and that these turbulent initial conditions significantly increase the accretion rates and allow the formation of high-mass stars within a couple of hundred thousand years \citep{mckee2002}. While the initial conditions are more turbulent than in low-mass star formation, other standard processes like disk formation or outflows can easily be part of this scenario. Central questions for this scenario are whether high-mass starless cores actually exist and whether the corresponding initial conditions are really more turbulent? We will discuss that in more detail in section \ref{turbulence}.

%\begin{marginnote}[]
%\entry{Clustering}{While many of the results for low-mass regions are based on isolated low-mass cores (e.g., \citealt{bergin2007}), the clustered properties of low-mass stars are also important (e.g., \citealt{lada2003,megeath2022}). In contrast, high-mass stars always form in clusters and hence multiple objects need to be taken into account (e.g., \citealt{beuther2007,zinnecker2007,tan2014}). Although one often compares low-mass star formation in nearby regions like Taurus or Perseus with high-mass star formation, the majority of low-mass stars form simultaneously with the high-mass stars in the clustered environment of high-mass star-forming regions.}
%\end{marginnote}

In parallel, the assumption of spherical symmetry was revisited in order to avoid isotropic radiation.  As soon as one drops that assumption and allows the formation of accretion disks, the radiation pressure and accretion problems can be overcome much more easily. Early work including rotational properties was conducted by \citet{jijina1996}, and \citet{yorke2002} calculated the rotating collapse in more detail. They also discuss the so-called flashlight effect where the radiation can easily escape perpendicular to the disk along the outflow cavities. These models were developed further in the coming years by, e.g., \citet{krumholz2005b,krumholz2007,vaidya2009,kuiper2010,kuiper2011,klassen2016,2016MNRAS.463.2553R,kuiper2018,rosen2022,oliva2023a,oliva2023b}, and by now there is ample observational evidence that accretion disks exist from low- to high-mass star formation (e.g., \citealt{andrews2018,beltran2016}). While ionizing radiation and the formation of hypercompact H{\sc ii} regions early in the evolution of high-mass stars may inhibit accretion (e.g., \citealt{churchwell2002}), already \citet{walmsley1995} suggested that high accretion rates may quench the development of the HCH{\sc ii} regions. This picture of trapped H{\sc ii} regions was then further developed in a series of papers \citep{keto2002,keto2003,keto2007}, and again accretion through equatorial disks are an important ingredient to make these processes happen (e.g., \citealt{keto2007,galvan-madrid2008,kuiper2018}, see sections \ref{disks} and \ref{feedback}).

Developing a scenario from even larger scales, \citet{vazquez2019} argue in the framework of a global hierarchical collapse of entire molecular clouds that star formation of all masses is largely driven by gravity. Whether clouds globally collapse is matter of debate since the seventies (e.g., \citealt{zueckerman1974a,zueckerman1974b,krumholz2014,evans2022}). \citet{vazquez2019} discuss that with the decrease of the Jeans mass with increasing density, the inner small-scale regions of giant molecular clouds collapse on much shorter timescales than the entire clouds. This then allows to also develop feedback processes on comparably short timescales and by that to disperse the lower-density cloud gas on the larger spatial scales.

\citet{padoan2020} developed a different model that also forms high-mass stars in the framework of converging gas flows. However, in contrast to the global hierarchical collapse that relies on the gravitational collapse of the clouds \citep{vazquez2019}, the inertial infall model builds on the idea of turbulent fragmentation of the molecular clouds \citep{padoan2020}. While conceptually different with gravity and turbulence at the origin of the collapse, an observational discrimination between these two approaches is extremely difficult and beyond the scope of this review.

Since star formation proceeds in a multi-fluid system of ionized and neutral gas components that can be described by magneto-hydrodynamic processes, gravity and radiation, the entire star formation picture can be described by many different processes and parameters. In the following, we will discuss and compare the important physical (and chemical) characteristics relevant for the star formation processes from low- to high-mass stars. While section \ref{properties} focuses on individual processes and their relevance for the different regime, section \ref{sythesize} aims at synthesizing that into a joint picture of star formation from low- to high-mass stars discussing the main similarities and differences.

\section{Properties and processes}
\label{properties}

\subsection{Environmental effects and bimodal star formation}
\label{environment}

\begin{figure}[ht]
\includegraphics[width=0.99\textwidth]{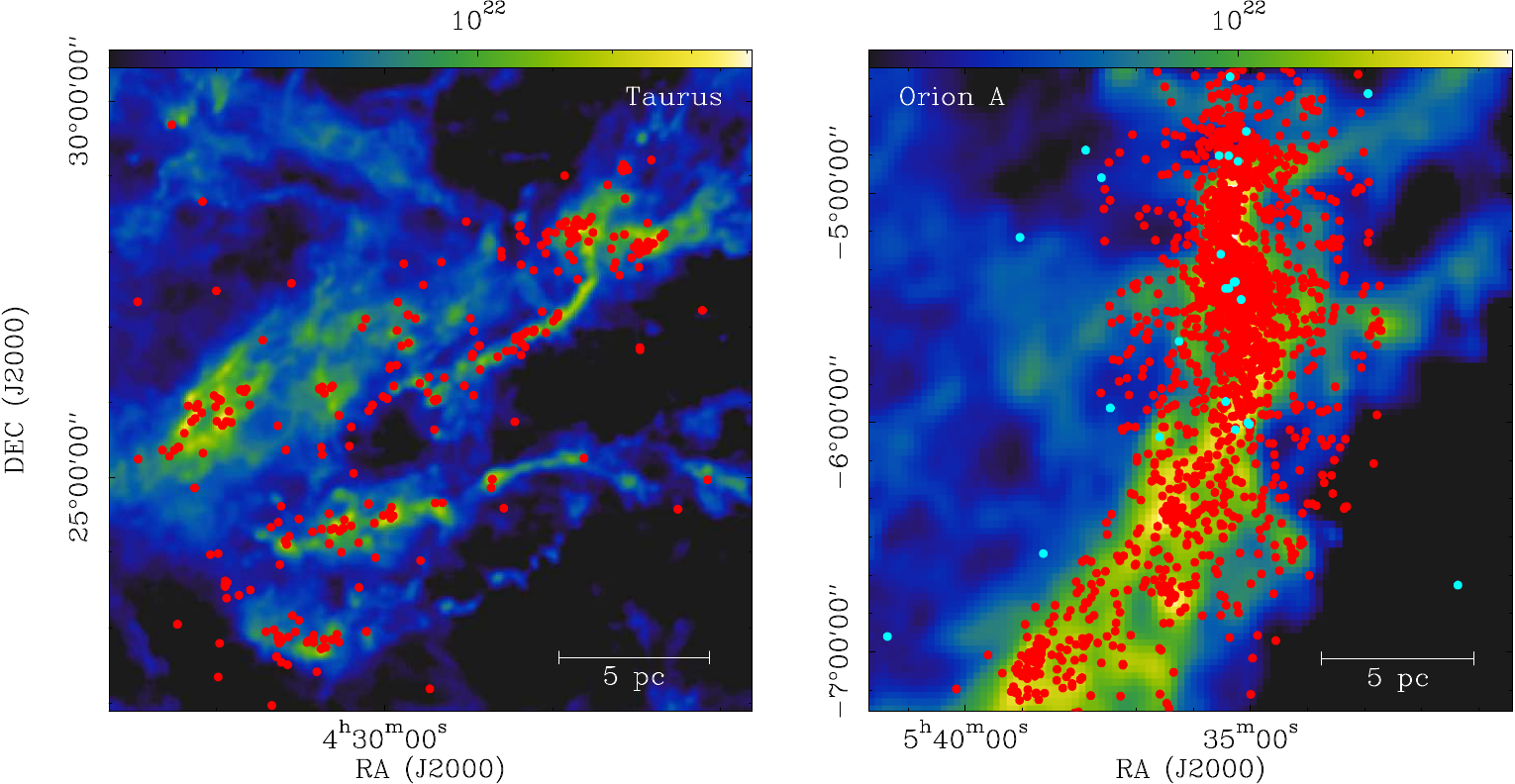}
\caption{Comparison between the Taurus and Orion A molecular clouds, the main
prototypes of the distributed and clustered modes of star formation. 
In both panels,
the background image shows the distribution of gas column density
determined from Planck observations \citep{planck2013XI}, and the red circles represent the 
positions of YSOs determined by \cite{luhman2018} for Taurus and \cite{megeath2012}
for Orion A (survey restricted to the high extinction part of the cloud).
The blue circles in the Orion A panel represent the positions
of stars B3 and earlier in the catalog of \cite{brown1994}. Both images
have been plotted 
at the same physical scale assuming Gaia distances of 141~pc
for Taurus and 432~pc for Orion \citep{zucker2019}.}
\label{fig_tau_vs_ori}
\end{figure}

\subsubsection{A diversity of star-forming environments}

The process of star formation can take place in a large variety of environments and give rise to a diversity of stellar densities. 
This is illustrated in
Fig.~\ref{fig_tau_vs_ori} with maps of the gas and young stars
in Taurus and Orion A, two of the most nearby examples of
the distributed and clustered modes of star formation.
The more sparse distribution of gas in Taurus has resulted in a median 
surface density of YSOs of about 2~pc$^{-2}$,
while the more concentrated distribution of mass in 
Orion A has produced a median stellar density more than one 
order of magnitude higher, and a local maximum that 
exceeds $10^4$~pc$^{-2}$ \citep{megeath2012}.

This diversity of star-formation environments is also correlated with a diversity
of stellar masses. Low-mass stars can be found in all known star-forming environments,
which range from isolated dark globules \citep{bok1978} to the densest parts of star clusters 
\citep{hillenbrand1998}. High-mass stars, on the other hand, seem to require
special conditions of formation, and most if not all of them seem to form
in clustered environments.
Whether it is possible to form a high-mas star in isolation is still an unsettled issue
 (e.g., \citealt{zinnecker2007}), 
but observations systematically show that isolated high-mass star formation
is at most a rare event. From a
study of 43 massive O-type field stars, \cite{dewit2005} found that after excluding 
runaway objects, only $4\pm2$\% of all O-type stars could be explained as having formed 
outside a cluster.
Even this small fraction of isolated high-mass stars could be an overestimate due 
to formation in undetected clusters \citep{stephens2017a}
and two-step ejections that cannot be traced back to the original
cluster \citep{pflamm2010}.
Since even the smallest star-forming regions contain enough material to form
a high-mass star, the main limitation to form high-mass stars in isolation seems not
to be mass availability, but the need for more extreme physical conditions. 

The difference in the environment in which low and high-mass stars can form
has been used to argue that star formation could be a bimodal process, meaning
``that the birth of low- and high-mass stars may involve separated mechanisms''
\citep{shu1987}. This idea has been proposed multiple times
over the past 60 years to explain different observational puzzles
\citep{herbig1962,eggen1976,elmegreen1978,guesten1982}.
Early models that invoked magnetic fields controlling the effects of gravity suggested that bimodality would arise as the difference between subcritical (strong) and supercritical (weak) fields in star-forming gas
(e.g., \citealt{shu1987,lizano1989}).
As a result of this view, the study of low-mass star formation has often been identified 
with the study of distributed environments like Taurus since these environments
are the ones that only produce low-mass stars.
Cloud-wide surveys of YSOs in nearby clouds, however, systematically show
that around 80\% of all stars (and therefore of all low-mass stars) 
form in embedded groups or clusters,
with only a small fraction of the star formation
taking place in a distributed manner (e.g., \citealt{carpenter2000,lada2003,evans2009}).
Large-scale infrared surveys, in addition, show that the
clustered and distributed modes
do not represent two distinct star-formation environments,
but are part of a continuum of stellar distributions that characterizes 
all star-forming clouds and that depends on the
density of the available gas (e.g., \citealt{bressert2010, megeath2022}).

\subsubsection{Is star formation bimodal?}
\label{bimodal}

Since low-mass stars can form both in isolation and in groups, a comparative
study of their properties in different environments provides a
test of the potential bimodality of star formation.
A useful property for this comparison 
is the shape of the 
initial mass function (IMF), which can be
systematically determined toward multiple
star-forming regions and used to search for
region-to-region variations. 
While current IMF determinations may still leave some room  
for variations under extreme conditions \citep{kroupa2001},
there is significant consensus that the shape of the IMF 
is nearly universal among galactic regions, with only possible differences 
at substellar masses (see \citealt{bastian2010} and \citealt{offner2014}
for reviews).
Even a persistent
discrepancy between the IMF of Taurus and of massive clusters like
the Orion Nebula Cluster, which 
suggested an stellar excess around 0.8~M$_\odot$ in Taurus
\citep{briceno2002}, has been
recently resolved when the new and more complete Gaia data have been used,
suggesting a lack of IMF variations across a range of stellar densities 
between 3-4 orders of magnitude \citep{luhman2018}.
This universality of the IMF, together with its simple 
form that includes a single power law connecting the low- and high-mass 
regimes, suggests that the formation of stars of 
different masses results from a continuous non bimodal
process.
 
Another property that could potentially reveal bimodality in star formation
is the multiplicity of the stars that it produces
(see reviews by \citealt{duchene2013} and \citealt{offner2023}, 
and Sect. \ref{fragmentation}
further discussion on multiplicity).
Main sequence stars are known to present multiplicity fractions that depend
strongly on the mass of the primary, with values
larger than 90\% for O-type stars, $\approx 50$\% for solar G-type stars, 
and $\approx 25$\%
for M-type stars (\citealt{sana2014}, \citealt{raghavan2010}, and \citealt{winters2019}, respectively). 
This characteristic mass dependence can be used 
to search for
environmental effects in the way stars are formed.
Comparing multiplicity fractions in different environments, 
however, is not straightforward. In contrast with the mass, the multiplicity
fraction of a star can change significantly during its lifetime 
as a result of interactions with nearby systems. For this reason, the
multiplicity fraction measured during the main sequence
does not necessarily reflect its value at birth.
Even at the T Tauri stage, the multiplicity fraction often presents signs of 
evolution, as in the case of the visual binaries of the Orion Nebula Cluster,
which are a factor of two less frequent than in other 
T Tauri associations, likely due to
interactions with other cluster members \citep{reipurth2007}.
Studies of the earlier protostellar phase may provide a better constraint on
the true multiplicity fraction at birth, but they are still very limited
due to the high extinctions. 
\cite{tobin2022}, for example, have used VLA and ALMA data to
compare the multiplicity fraction of protostars in 
the Perseus and Orion clouds, and found no significant evidence for 
differences between the two. While the current population of protostars in 
the Orion cloud studied by these authors
is more distributed than the one responsible for the Orion 
Nebula Cluster, the above comparison provides a first hint that the
multiplicity fraction of protostars may only depend weakly on the 
environment. Further work on the multiplicity fraction of embedded 
protoclusters is clearly needed to confirm this result (see also Sect.~\ref{fragmentation}).

In the absence of more definitive evidence, the natural conclusion is that
environmental effects play only little role in the formation of low-mass stars,
and that star-formation is therefore not strictly bimodal. 
High-mass star formation does require larger masses of dense gas than low-mass star formation, 
but since high-mass stars form in clusters, the same global 
conditions that produce high-mass stars
can also produce multiple low-mass stars with similar properties and in a similar 
proportion to less massive regions.
Studies of star
formation in different environments can therefore be seen as providing 
complementary information on the physics of stellar birth.
Studies of
nearby regions of distributed star formation allow us to 
zoom in on the formation of individual low-mass stars at the expense of missing
their high-mass counterparts, while the study of more distant 
high-mass regions allow us to investigate the formation of the
high-mass end of the stellar population
at the expense of a loss in angular resolution. 

\subsection{Infall motions}
\label{infall}

\subsubsection{The spectroscopic signature of infall and its searches}

Gravitational infall represents the ultimate cause of stellar birth, so finding 
evidence for infall motions toward embedded protostars has been a major 
goal of both low- and high-mass star-formation studies
(see \citealt{evans2003} for a historical account). 
The preferred tool to search for infall has been the analysis of optically
thick molecular emission. If a contracting cloud is 
spherical and has a density or temperature profile that decreases outwards, the 
excitation of any molecule will also decrease toward the outer layers. 
Under any realistic velocity field, the redshifted 
part of the radiation coming from the cloud interior will be absorbed 
by the outer layers, causing a characteristic spectral feature  that is
commonly referred to as the infall asymmetry
(e.g., \citealt{leung1977,zhou1994,myers1996}). To
unambiguously identify this feature, it has become customary to require that the
detection of an asymmetric optically thick line is complemented with the detection
of a symmetric line in an optically thin tracer that
guarantees the
absence of additional cloud components or any other type of asymmetry
in the cloud velocity field (e.g., \citealt{myers2000}).

The first systematic searches for infall toward low-mass protostars were motivated
by the detection of infall signatures at the
center of the B335 globule \citep{zhou1993}. These searches used optically
thick lines of H$_2$CO and CS, and found significant excess of sources 
with infall asymmetry, especially in samples of Class 0 protostars
\citep{gregersen1997, mardones1997}.
To quantify the inward motions, \cite{mardones1997} defined the $\delta V$ parameter 
as the difference between the peak velocities of a thick and a thin line 
normalized by the
width of the thin line 
($\delta V = (V_{\mathrm{thick}}-V_{\mathrm{thin}})/\Delta V_{\mathrm{thin}}$), 
so a negative $\delta V$ value indicates infall asymmetry. 
\cite{mardones1997} considered this parameter as significant if its absolute 
value exceeded 0.25, and using this criterion, they estimated an excess
of inward motions 
of $\approx 25$\% from their survey.
They also found that in some sources the sign of $\delta V$ changes with 
tracer, indicating possible contamination by outflows, or
that the contraction motions  
may be more complex than predicted by a simple spherical model. 
These initial findings have been confirmed by later infall 
studies using a variety 
of optically thick tracers (e.g., \citealt{takakuwa2007}, \citealt{mottram2017}).

Infall signatures have also been detected toward starless cores \citep{tafalla1998, caselli2002}, and a systematic search toward 220 of them by 
\cite{lee1999} and  \cite{lee2001} found a $\approx 20$\% excess of blue asymmetries, again suggestive of a prevalence of inward motions. Since the targets are starless, the motions have been interpreted not as protostellar-forming collapse but as arising from large-scale core-forming contraction flows. 

Multiple infall searches have been carried out toward regions with high-mass
star formation. 
\cite{wu2003} observed 28 massive clumps associated with H$_2$O masers and found a 
$\approx 30$\% excess of significant infall profiles, while \cite{fuller2005} carried out a similar study of 77 high-mass sources using several transitions of HCO$^+$, and found a fraction of blue profiles similar to that of \cite{wu2003} in the low-J transition of HCO$^+$, but no excess in the high-J lines. \cite{fuller2005} interpreted this result as an indication of the infall motions being stronger in the low-density gas outer layers that are responsible for the absorption of the low-J transitions.  Additional infall searches have increased the variety of tracers and physical conditions explored, and in most cases have confirmed the small 
but significant  prevalence of inward motions (e.g., 
\citealt{wu2007}, \citealt{sun2009}, \citealt{chen2010}, 
\citealt{wyrowski2012,wyrowski2016}). 
The largest search toward high-mass regions so far has been carried out by \citet{jackson2019}, 
who observed the 3\,mm lines of HCO$^+$ and 
N$_2$H$^+$ toward 1093 condensations from the ATLASGAL 850\,$\mu$m survey. 
In addition to finding a small excess of infall profiles in their sample, \cite{jackson2019} found that the excess of infall profiles decreases from the earliest evolutionary stages (from quiescent cores to ultracompact H{\sc ii} regions) to the more evolved phases (H{\sc ii} and photodissociation regions), suggesting the existence of an evolutionary trend. 

\subsubsection{Low-mass versus high-mass results and comparison with models}

\begin{figure}[ht]
\includegraphics[width=0.99\textwidth]{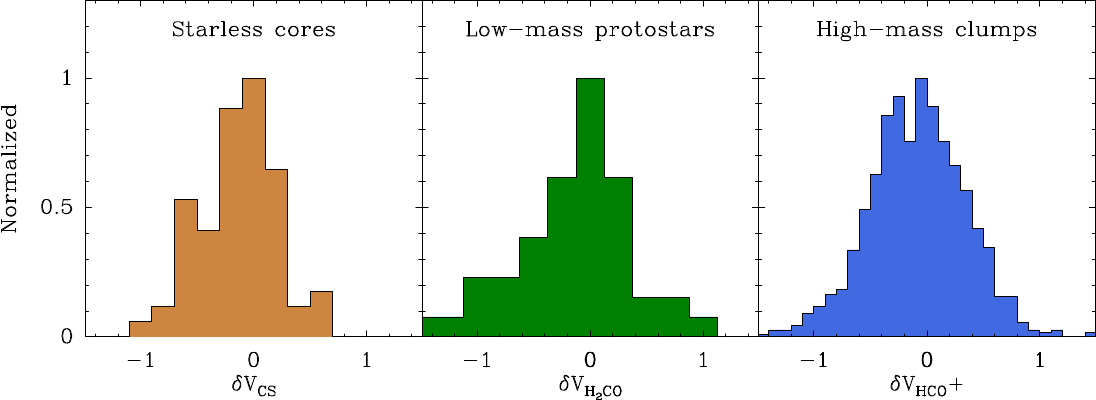}
\caption{Normalized histograms of the $\delta V$ parameter derived from 
infall searches towards low-mass 
starless cores (left panel, \citealt{lee2001}), low-mass
protostars (middle panel, \citealt{mardones1997}), and high-mass clumps 
(right panel, \citealt{jackson2019}). All $\delta V$ estimates use 
N$_2$H$^+$ as the thin tracer and the thick tracer indicated in the 
abscissa label. Despite the large differences between their targets, the three
histograms present similar $\delta V$ distributions in terms of shape, width,
and slight excess of negative values (indicative of contraction motions).}
\label{fig_deltav}
\end{figure}

While no systematic comparison between infall searches toward low- and high-mass star-forming regions has been carried out, a simple inspection of the 
published results shows strong similarities between the results from
both types of regions in 
terms of the $\delta V$ distribution. This is illustrated in
Fig.~\ref{fig_deltav} 
with a comparison between the $\delta V$ distributions for starless cores, 
low-mass protostars, and high-mass clumps.
As the figure shows, all $\delta V$ distributions 
present a similar well-defined maximum near zero and 
an approximate span from -1 to 1. Indeed, 
\cite{jackson2019} found that the $\delta V$ distribution 
of their sample of high-mass condensations had the same FWHM of about 1 as the 
distribution of low-mass protostars from \cite{mardones1997} despite the 
width of the spectra 
in the two samples differing by a factor of about 3. This predominance of 
$\delta V$ values significantly lower than 1 indicates that the inward motions 
seen in both low- and high-mass star-forming regions must be significantly 
smaller than the FWHM linewidth of an optically thin line, and therefore smaller than
the free-fall velocity, a result previously emphasized for high-mass 
star-forming regions by \cite{rolffs2011} and \cite{wyrowski2016}.

Although the
$\delta V$ parameter provides a simple test for the presence of infall 
asymmetries, estimating the speed of the inward motions requires detailed 
modeling of the spectra. Given the high optical depth of the lines used to 
search for infall, their radiative transfer is highly sensitive to the geometry 
and velocity field of the source, which are often poorly constrained. 
As a result, detailed modeling of infall line profiles has been restricted to a 
few simple objects, such as B335 and BHR71, and usually assuming some theoretical 
prediction of the infall conditions \citep{zhou1993,yang2020, evans2023}.
A more common approach to estimate inward velocities has been to derive a single 
characteristic value using a simple analytic model of a collapsing cloud, as the 
one presented by \cite{myers1996}, which assumes that the cloud consists of two 
parallel gas layers approaching each other (see also \citealt{devries2005}). 
Using this model, \cite{lee2001} derived inward velocities between 0.05 and 
0.09~km~s$^{-1}$ for the best infall candidates of their starless-cores survey. 
For low-mass protostars, no systematic analysis of the inward velocity has 
been presented, but the modeling of individual sources indicates that the speeds 
typically range from less than 0.1~km~s$^{-1}$ to several tenths of km~s$^{-1}$ 
in the most extreme cases, such as L1251B (0.35~km~s$^{-1}$, 
\citealt{myers1996}) and NGC1333 IRAS4A (0.68~km~s$^{-1}$ 
\citealt{difrancesco2001}). For the high-mass 
regions, \cite{fuller2005} derived velocities between ~0.1 and 
 1~km~s$^{-1}$, and quoted 0.2~km~s$^{-1}$ as a “typical” value, while 
\cite{wyrowski2016} 
derived inward velocities between 0.3 and 2.8~km~s$^{-1}$.
All these values indicate 
a systematic trend of increasing inward velocity toward the more massive 
regions, as would be expected from their deeper potential wells.

The above estimates provide useful constraints to 
models of star formation, but most values quoted so far likely trace the upper 
envelope of the distribution of inward velocities. Infall studies have mostly 
focused on sources with prominent infall asymmetry, so the published results 
tend to represent the best-case scenario of infall evidence. As illustrated in 
Fig.~\ref{fig_deltav}, unbiased surveys show that the majority of the $\delta V$ estimates lie in the vicinity of zero, and must therefore correspond to small 
values of inward and even outward velocity. 
Since in some cases these values have been measured toward 
Class 0 protostars, which are believed to be actively 
accreting (e.g., \citealt{andre1993}), 
the spectroscopic infall signature seems to be at best a weak indicator of the 
true star-forming motions. This low sensitivity of the infall signature likely 
results from the complexity of the infall motions that take place in real 
star-forming regions, which deviate from the spherically-symmetric pattern
assumed by the standard infall profile analysis. 

\cite{smith2012,smith2013} have investigated this problem using
a numerical simulation of a turbulent collapsing cloud coupled with a radiative 
transfer model to predict the emergent spectra of optically thick and thin line 
pairs under realistic infall conditions.
In this simulation, the gravitational motions deviate strongly from spherical 
symmetry and give rise to prominent filamentary structures like those 
seen in real star-forming clouds (e.g., \citealt{andre2010}, 
\citealt{molinari2010}, \citealt{hacar2023}, \citealt{pineda2023}). Since filament formation precedes core formation, the very same motions that create the pre-stellar cores also surround them with the anisotropic velocity field responsible for the filament formation. As a result, the optically thick emission from a core region contains a large contribution from the surrounding filamentary gas due to a combination of high optical depth and molecular freeze out at high density. 
In low-mass star-forming regions, 
\cite{smith2012} found that this extended contribution is able to
hide the expected blue-asymmetric profile in
more than 50\% of the cases despite the region being in a true state of gravitational infall. It can even reverse the infall profile if an asymmetry in the 
surrounding filament gives rise to a one-sided accretion flow onto the core that 
comes from behind.
A similar decrease in the frequency of observable infall signatures is predicted to occur in regions forming high-mass stars, for which \cite{smith2013} found that the optically thin tracers present multiple velocity components in their spectra, and that the thick lines only display marginal blue asymmetries in most cases. The classical infall signature, therefore, seems strongly sensitive to the overall geometry of the cloud, and may only trace
infall reliably under very favorable symmetric conditions.

\subsubsection{Filaments: beyond spherical infall}

If star-forming collapse deviates from spherical symmetry, characterizing it
requires a more detailed determination of the gas kinematics than the
line-of-sight information provided by the spectral asymmetry. 
Such a determination is difficult for a general three-dimensional geometry, 
but the prevalence of filamentary structure in the star-forming gas offers
additional constraints under favorable conditions. In a study of the L1517 dark cloud, \cite{hacar2011} found that both the velocity field and the column density along some of the filaments presented a similar pattern of wave-like 
oscillations with a relative offset of about one quarter of the wavelength. This 
pattern matches the expected signature of gas undergoing filamentary fragmentation \citep{gehman1996},
and suggests that the observed motions in L1517 may represent core contraction along the filaments. Similar wave-like patterns in the velocity field have been identified in regions of both low- and high-mass star formation,  although the one quarter of
the wavelength offset with the column density profile seems an elusive feature 
(e.g., \citealt{hacar2013}, \citealt{henshaw2014}, \citealt{tackenberg2014},
\citealt{zhang2015}, \citealt{liu2019}, \citealt{shimajiri2023}, 
\citealt{yoo2023}). 
A related motion identified along some filaments is the streaming of material toward a massive condensation in a hub-filament region \citep{myers2009}. Examples of these motions have been found by \cite{kirk2013} in the Serpens South protocluster and \cite{peretto2014} in the SDC13 IRDC, and illustrate that even large-scale collapsing motions in clouds can deviate strongly from an ideal spherical symmetric configuration. In this context, \citet{kumar2020} discussed a picture where filaments form early, associated already with some low-mass star formation. During the ongoing collapse within the region, some of the filaments may merge resulting in hub-filament structures, where high-mass star formation may later occur (see \citealt{motte2018} for a similar view).

At smaller spatial scales, there is also increasing evidence that the collapse toward low-mass protostars is highly anisotropic (e.g., \citealt{tobin2024}). Extinction maps of the vicinity of Class 0 protostars derived from Spitzer Space Telescope data indicate the prevalence of highly anisotropic distributions of matter on scales larger than 1,000 au \citep{tobin2010}, and ALMA polarization observations of the submillimeter dust continuum have revealed a possible streamer of accreting material in  Serpens Emb 8(N) \citep{legouellec2019}. In addition, high-angular-resolution molecular line observations and simulations have started to show asymmetric distributions of gas that connect the core-envelope environment with the central protostellar disk and have kinematics indicative of streaming infalling motions \citep{yen2014,pineda2020,cabedo2021,valdivia-mena2022,pineda2023,kueffmeier2024}. Similarly, high-resolution studies of high-mass regions also revealed filamentary structures that feed the central massive hubs or cores (e.g., \citealt{peretto2014,trevino-morales2019,kumar2022,li2022,wells2024}).

Different signatures of ongoing infall motions have recently been reported by high-resolution ALMA observations. Based on the ALMA ASHES survey of very young high-mass regions, \citet{morii2023} do not find individual sub-cores in any of the regions that contain enough mass to form a massive star. Therefore, they argue that the cores need to continuously accrete material from the massive clump structure to form a high-mass star in the end. In the large ALMAGAL survey of more than 1000 high-mass star-forming regions of all evolutionary stages, Coletta et al.~(subm.) find that the most massive core per region also increases in mass with evolutionary stage (see also \citealt{morii2024}). In another ALMAGAL study, \citet{wells2024} recently found that the flow rates towards the central cores increase with core mass roughly like $\dot{M}\propto M^{0.68}$. The latter relation is consistent with a model of tidal lobe accretion, in which the potential of the forming cluster is still dominated by the gas component \citep{clark2021}.

Further work is needed to characterize the motions in the vicinity of protostars of different masses, but the available data already suggest that non-spherical infall is the norm in star-forming regions, and that understanding these motions is a necessary step to finally connect the flow of matter from cloud scales to the protostar-disk systems, where rotation dominates the gas kinematics (for disk formation, see Sect. \ref{disks}).

\subsection{Density structure}
\label{density_structure}

\subsubsection{Density slopes}
\label{density}

The density distributions $\rho(r)$ are a critical parameter characterizing the structures of star-forming regions. Classical stability calculations of the Bonnor-Ebert sphere result in density distributions $\rho\sim r^{-p}$ with the power-law exponent $p\sim 2.0$ at large spatial scales \citep{ebert1955,bonnor1956}. Early numerical work of the collapse of a cloud with initially uniform density distribution lead to a distribution where $p$ ranges between $\sim$1.7 and $\sim$2 (\citealt{penston1969,larson1969}, see also \citealt{hunter1977,whitworth1985}). The collapse solution of the singular isothermal sphere by \citet{shu1977} results in an outer almost static envelope with $p\approx 2$ whereas the inner collapsing part approaches $p\approx 1.5$. Early theories of high-mass star formation proposed a different, logathropic equation of state that resulted in flatter density distributions with $p\approx 1$ \citep{mclaughlin1996,osorio1999}. 

%\begin{figure}[ht]
\begin{wrapfigure}{r}{0.6\textwidth}
\includegraphics[width=0.6\textwidth]{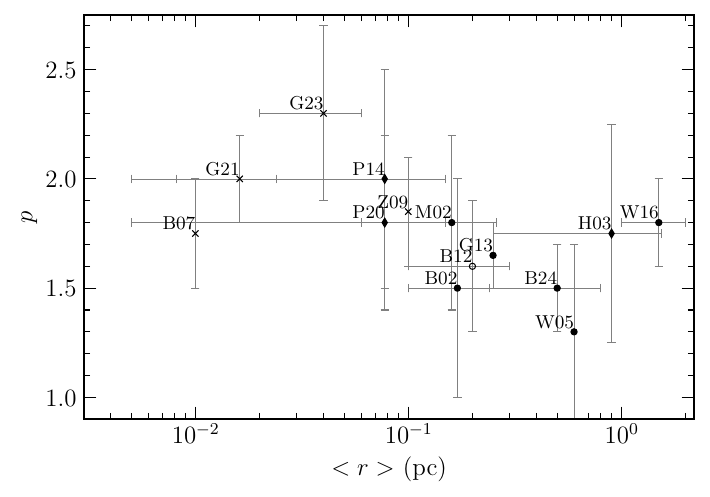}
\caption{Comparison of density index $p$ plotted versus the corresponding typical core and clump sizes $<r>$. This is an update of Fig.~10 from \citet{gieser2021}. The labels correspond to the references listed in \citet{gieser2021} plus a few new references: G21 \citep{gieser2021}, G23 \citep{gieser2023} and B24 \citep{beuther2024}. Crosses and diamonds mark interferometer and single-dish data, respectively.}
\label{p_index}
\end{wrapfigure}
%\end{figure}

These theoretical predictions have been observationally tested. Observational studies of cores in low-mass star-forming regions typically infered density distributions with $p$ ranging between 1.5 and 2 (e.g., \citealt{motte2001,alves2001}). For comparison, in the high-mass regime, studies of the density distributions of large-scale parental gas clumps tracing the cluster-forming regions ($\sim$1pc scale) resulted typically in density distributions with $p$ between 1.5 and 2 as well (e.g., \citealt{beuther2002a,mueller2002,hatchell2003,williams2005,palau2014,wyrowski2016}). \citet{giannetti2013} found a tentative difference in $p$ of 1.5--1.8 for active star-forming clumps and slightly lower values of $p$ around 1.2--1.4 for earlier more quiescent regions. Potentially flatter density distributions at earlier evolutionary stages were also suggested by \citet{beuther2002a} or \citet{butler2012}. Figure \ref{p_index} presents a literature compilation of derived density indices $p$ versus corresponding typical sizes $<r>$.

On smaller scales, early interferometric studies of only a few cores in high-mass regions with the Plateau de Bure Interferometer (PdBI) and the Submillimeter Array (SMA) also found density power-law distributions between 1.5 and 2.0 \citep{beuther2007b,zhang2009}. More recent studies of larger samples with the Northern Extended Millimeter Array (NOEMA, formerly PdBI) of the density distribution of small-scale cores in high-mass star-forming regions found power-law indices of typical high-mass protostellar objects (HMPOs) with $p\approx 2.0\pm 0.2$ whereas younger regions appeared slightly shallower with $p\approx 1.6\pm 0.4$ \citep{gieser2022}. Combining the previous with additional data from the Atacama Large Millimeter Array (ALMA), \citet{gieser2023} compiled the small-scale core density power-law distributions with a clustering of $p$ around 2, but also exhibiting a comparably large scatter roughly between 1.0 and 2.5.

A recent comparison of large clump-scale versus small core-scale density distributions for a sample of HMPOs revealed a steeping of the density profiles with values clustering around 1.5 for the larger clump scales and values around 2.0 for the smaller core scales \citep{beuther2024}. This is consistent with the collapsing cores being embedded in larger-scale clouds that typically have shallower density distributions. Analytic and numerical modeling results in similar density structures where the $p\approx 2$ profile is considered an "attractor", i.e., initially flatter density distributions typically approach the $p\approx 2$ slopes (e.g., \citealt{naranjo-romero2015,gomez2021}). 

In summary, observational studies are generally consistent with theoretical predictions. Density power-law distributions with a power-law index $p$ typically between 1.5 and 2.0 are found for the majority of star-forming regions, independent of mass. These density distributions are typically embedded in larger-scale parental molecular clouds that may exhibit even flatter density profiles. This is also consistent with the tentative finding of a few studies of potentially flatter density distributions at earlier evolutionary stages. 

\subsubsection{Column density and density thresholds for star formation}
\label{thresholds}

In addition to the density slopes, variations in the absolute densities within star-forming regions may also contribute to differences in the mode of star formation. Early analysis of the column density distributions in molecular clouds revealed a deviation from pure lognormal probability density distributions (PDFs) at high column densities (e.g., \citealt{lombardi2006,lombardi2008,kainulainen2009}). This deviation can be interpreted as a column density threshold above which star formation proceeds (e.g., $A_v>2-5$\,mag or $A_v>8$\,mag in \citealt{kainulainen2009} or \citealt{lada2010,heiderman2010,evans2014}, respectively). In the high-mass regime, \citet{krumholz2008} suggested a column density threshold of 1\,g\,cm$^{-2}$, corresponding to H$_2$ number column densities of $\sim$3$\times 10^{23}$\,cm$^{-2}$ or a visual extinction of $\sim$320\,mag (e.g., \citealt{frerking1982}). In a series of papers, the mass--size relation $m(r)$ was established as a different threshold where the dividing line between low- and high-mass star formation is roughly $m(r)\approx$870\,M$_{\odot}(r/{\rm pc})^{1.33}$ \citep{kauffmann2010a,kauffmann2010b,kauffmann2010c}, corresponding to a H$_2$ column density threshold of $\sim$5.4$\times 10^{22}$\,cm$^{-2}$ or a visual extinction of $\sim$58\,mag. This latter empirical column density threshold for high-mass star formation is a factor of a few lower than the theoretical threshold proposed by \citet{krumholz2008}, but it is almost a factor 10 higher than the previous estimates for low-mass regions (e.g., \citealt{kauffmann2010a}).

In addition to the column densities, one can also estimate volume density thresholds for star formation. Assuming spherical symmetry and cloud radii of 1\,pc,
for low-mass regions, the mean volume densities corresponding to the $A_v$ range of 2-8\,mag (see previous paragraph) are roughly between $1.5\times 10^2$ and $5.8\times 10^2$\,cm$^{-3}$. For comparison, the corresponding volume mean densities from the high-mass thresholds by \citet{krumholz2008} and \citet{kauffmann2010c} are $\sim 2.3\times 10^4$\,cm$^{-3}$ and $\sim4.2\times 10^3$\,cm$^{-3}$, respectively. Hence, also the mean required volume densities in the high-mass regime are roughly an order of magnitude higher than in low-mass regions.

Alternative values have been estimated by
\citet{lada2010} and \citet{evans2014}, who inferred that the volume density associated to the $A_v=8$\,mag extinction threshold for low-mass star formation corresponds to roughly $10^4$ and $6\pm4 \times 10^3$\,cm$^{-3}$, respectively. \citet{kainulainen2014} conducted a 3D decomposition of a sample of low-mass star-forming regions and estimated a volume density star-formation threshold of $5\times 10^3$\,cm$^{-3}$. 

To give a few more examples, assuming again spherical symmetry and using the masses and sizes reported in \citet{krumholz2005} for a few low- and high-mass regions, one can also estimate their mean densities, and one gets mean volume densities of the low- and high-mass regions of $\sim$2.7$\times 10^2$\,cm$^{-3}$ and $\sim$1.8$\times 10^6$\,cm$^{-3}$, respectively. In a different study, based on example regions for a low-, intermediate- and high-mass star-forming region, \citet{beuther2013} inferred mean densities  of the corresponding cores. Based on the Herschel far-infrared data presented by \citet{nielbock2012}, they estimate a mean density of $1.2\times 10^4$\,cm$^{-3}$ over a diameter of 48.000\,au for the low-mass B68 region. For the intermediate-mass region IRDC\,19175, \citet{beuther2013} estimate a mean density of $2.6\times 10^4$\,cm$^{-3}$ over a diameter of 45.000\,au, and for the high-mass region IRDC\,18310 the estimate results in mean densities of $1.5\times 10^6$\,cm$^{-3}$ over a slightly smaller diameter of 18.000\,au (see also section \ref{fragmentation} and Fig.~\ref{jeans-length}). 

In summary, while there are quantitative differences in the predictions of column density and volume density thresholds for star formation, the studies agree that column and volume densities are significantly higher (roughly an order of magnitude) in high-mass regions compared to their low-mass counterparts. It is still not settled whether column or volume density is the more important parameter to establish star formation thresholds. While theorists suggest that volume densities should be the controlling factor (e.g., \citealt{krumholz2012,padoan2014}), others argue that the column densities may be the more important constraint (e.g., \citealt{mckee1989,evans2014}).

\subsection{Turbulent and thermal support}
\label{turbulence}

One assumption often made is that high-mass star-forming regions are more turbulent than their low-mass counterparts (e.g., \citealt{mckee2003}). This assumption was initially based on observations of high-mass star-forming regions that already contained ongoing star formation processes (e.g., \citealt{plume1997}). Hence, the observed velocity dispersion did not necessarily represent the initial conditions. With the arrival of mid-infrared Galactic plane surveys like MSX and Spitzer, catalogs of infrared dark clouds (IRDCs) as candidates for the initial conditions emerged (e.g., \citealt{egan1998,peretto2009}). These IRDCs have been studied intensely since then with single-dish and interferometric observations. Generally speaking, most studies seem to agree that the velocity dispersion at these early evolutionary stages is typically very low, especially when analyzing high-spatial resolution interferometer data resolving the sub-structures (e.g., \citealt{pillai2011,sanchez-monge2013,beuther2015,morii2021,li2022,li2023,wang2024,zhang2024}, see also section \ref{fragmentation}). 

More specifically, e.g., \citet{beuther2015} reveal several narrow velocity components ($\Delta v\sim$0.3\,km\,s$^{-1}$ at 0.2\,km\,s$^{-1}$ resolution) towards individual cores within the high-mass starless core candidate IRDC\,18310-4. Only the overlap of these multiple components in lower-angular-resolution studies merge these individual narrow components into apparent broader lines that may mimic larger turbulence ($\Delta v\sim$1.7\,km\,s$^{-1}$ in single-dish data, \citealt{sridharan2005}, see also \citealt{smith2013} for corresponding simulations). For comparison, analysing ALMA data of the young high-mass filamentary region NGC6334S, \citet{li2022} find that the gas filaments are largely supported by thermal motions where the non-thermal motions are predominantly subsonic or transonic. In a recent ALMA study of the ASHES sample of 70\,$\mu$m dark high-mass regions, \citet{li2023} also find that the virial parameter decreases with core mass (see also \citealt{kauffmann2013} or \citealt{friesen2024}).

In summary, the studies appear to converge to a picture where the turbulent support in high-mass star-forming regions does not differ significantly from that in low-mass regions.

\subsection{Magnetic fields}
\label{magnetic}

The influence of magnetic fields on star formation processes is an important topic, but often the observational constraints are not as conclusive as one may have hoped for (see some recent reviews \citealt{crutcher2012,li2014,hull2019,pattle2019,pattle2023,liu2022}). Theoretical magneto-hydrodynamical (MHD) modeling predicts that fragmentation is reduced by strong magnetic fields (e.g., \citealt{tomisaka2002,myers2013,li2014b,commercon2011,commercon2022,matsushita2017,hennebelle2008,hennebelle2022}). 

\citet{li2014} report that magnetic fields have the tendency to preserve their orientation from large cloud- to small core-scales. On large spatial scales, data from the Planck mission revealed that at low column densities the magnetic field largely follows the gas structure, but going to higher column densities (a few times $10^{21}$\,cm$^{-2}$) magnetic field and denser filaments are typically oriented perpendicular to each other (e.g., \citealt{planckXXXV,fissel2019,soler2019}). Magnetic field studies of individual star-forming regions revealed further structural changes of the magnetic field morphology, for example, gas flowing along filaments can again align the magnetic field with the filamentary structure (e.g., \citealt{koch2014,pillai2020,beuther2020,stephens2022}, see also simulations, e.g., \citealt{klassen2017,gomez2018}). Recently, \citet{beltran2024} report self-similar hourglass-shape morphological structures in the high-mass hot core G31.41 where the collapse in the outer parts of the core are slightly sub-Alfvenic and become super-Alfvenic close to the center. This implies slower collapse in the outer regions, potentially along magnetic field lines, whereas closer to the center the collapse accelerates and overwhelms the magnetic field pressure. On the smallest disk-scales, the magnetic field is expected to attain even a more rotational configuration following the disk kinematics (e.g., \citealt{seifried2015,beuther2020,sanhueza2021}). 

\citet{pattle2023} review the important magnetic field density relation ($B\sim n^\kappa$). While collapsing clouds with flux-freezing produce a $\kappa\approx 2/3$ \citep{mestel1966}, ambipolar diffusion models find values of $\kappa$ between 0.0 and 0.5 (e.g., \citealt{mouschovias1999}). In a compilation of data, \citet{crutcher2010} inferred a $\kappa\sim 0.65$ but re-analysis of the same data by \citet{tritsis2015} resulted in $\kappa$ being consistent with $\sim$0.5. Re-analysis of all dust polarization data from the literature finds a $\kappa\sim 0.57$ \citep{liu2022b}, whereas \citet{whitworth2024} find slopes of $\kappa\sim 0.27$ and $\sim$0.54 for densities smaller and larger than $\sim$924\,cm$^{-3}$. It turns out that the uncertainties in the observational determination of the magnetic field and density are too high to more accurately determine $\kappa$ (e.g., \citealt{jiang2020,liu2022b}). Furthermore, the magnetic virial parameter decreases with increasing column density, indicating that the magnetic support decreases with (column) density, allowing the dense cores to collapse and form stars (e.g., \citealt{liu2022b}).

Although only few polarization observations of high-mass IRDCs exist, there seems to be a consistent picture of ordered magnetic fields perpendicular to the main axis of the filaments, consistent with simulations that have no strongly super-Alfvenic turbulence  (e.g., \citealt{pillai2015,falceta-goncalves2008,liu2018,tang2019,liu2024}). \citet{pillai2015} argue that higher initial magnetic field strengths are needed in high-mass star formation to enable the formation of dense, massive and large filamentary structures. For the IRDC G28.37, also discussed with respect to the magnetic field orientation by \citet{liu2024}, \citet{kong2019} find that most outflows are orientated almost perpendicular to the main filament axis (see also Fig.~\ref{g28_outflows} related to the outflow section \ref{outflows}), which may stem from continuous mass transport from the filament to the embedded cores. This could then well be related again to the magnetic field structure (e.g., \citealt{liu2024}). In comparison to that, such an outflow-filament (mis-)alignment has never been reported in low-mass regions (e.g., \citealt{stephens2017b}). In addition, \citet{pattle2023} suggest that low-mass star formation may proceed rather in environments around magnetic criticality whereas high-mass star formation may occur in more supercritical environments. The latter would imply that in high-mass star formation the magnetic fields were less capable to prevent or slow down the collapse. However, this difference is not tightly confirmed yet. For possible relations between the magnetic field and fragmentation, see the following Sect.~\ref{fragmentation}.

In summary, while there are indications that the magnetic field may play a stronger role on cloud- and filament-formation scales in the high-mass regime, on small core-scales, the larger gravitational potential of high-mass regions may then downweight the magnetic field importance compared to the low-mass regions. All these spatial regimes certainly need further investigations. 

\subsection{Fragmentation and multiplicity}
\label{fragmentation}

Since stars of all masses dominantly form in a clustered mode, understanding the initial fragmentation processes of the parental gas clumps is crucial for a thorough understanding of star formation. One way to characterize cluster properties is the finding of an almost universal initial mass function (IMF), starting with the seminal work by \citet{salpeter1955}. In this review, we discuss the IMF only in the context of the environment (Sect.~\ref{bimodal}) and refer for more detailed discussions to selected critical papers and reviews (e.g., \citealt{salpeter1955,miller1979,kroupa2002,chabrier2003,corbelli2005,bastian2010,krumholz2014,offner2014,hennebelle2020,2024ARA&A..62...63H}). 

With respect to multiplicity, a few things are independent of stellar mass, while the processes leading to close separation binaries clearly differ between low- and high-mass stars. Common among all type of stars is, e.g., a rather indistinguishable slope of the companion fraction as function of binary separation $a$ for wide-separation binaries ($a > 30\mbox{ au}$) \citep{offner2023} and the mass ratio $q$ of these binaries match random pairings drawn from a Salpeter IMF down to $q \approx 0.4 \ldots 0.3$, for $q < 0.3$ it is slightly top-heavy \citep{2017ApJS..230...15M}. This seems to point to a picture where star formation can be described as a process continuous on the mass scale without specific processes distinguishing high-mass stars from the low-mass range. \citet{offner2023} describe several mechanisms responsible for the fragmentation, including filament fragmentation, core fragmentation, disk fragmentation and capture. While the first two act on scales of star-forming regions (pc to sub-pc), disk fragmentation is typically important on much smaller scales below 1000\,au. 
The formation of multiple objects by capture needs close passages of objects and is closely related to the dynamics of the forming systems that can happen on scales of entire clusters.
The simulation study by \citet{2023A&A...674A.196K} supports the idea that the fractions of systems that form via dynamical capture and core fragmentation are rather independent on the initial gas density although these simulations focus solely on low-mass star formation.

The most striking differences in low- and high-mass multiplicity are related to the number of companions (Sect.~\ref{bimodal}) and the fraction of close binaries. O-type stars have a second peak of the companion fraction at small binary separations $a$, and B-type stars have at least a shallower drop of the companion fraction as function of separation toward smaller separations than all other low-mass stars; also the value of the mean separation of inner binaries is smaller for high-mass stars than for solar-type stars \citep{offner2023}. Close-separation binaries ($a < 0.5 \mbox{ au}$) do not favor any specific mass ratio $q$ (with a minor $10\%$ of twin excess, \citealt{2017ApJS..230...15M}). While solar-type stars show a clear metallicity dependence (close-separation binary fraction increases toward lower metallicity), for high-mass stars no statistically significant trends with metallicity can be observed \citep{2013ApJ...778...95M}. In short, the main distinctive difference is that high-mass stars dominantly form as multiples and favor close-in companions. All of the individual indices above suggest different formation physics of close multiples for high-mass stars. This is in line with disks around forming high-mass stars being more prone to fragmentation (e.g., \citealt{kratter2008,kratter2016,2020A&A...644A..41O}). When the probability to fragment reaches nearly $100\%$ for present-day stars, the process also becomes independent of metallicity (see also Sect.~\ref{disks} for more discussion on disk fragmentation). Binary formation via disk fragmentation can also increase the spin of the primary star \citep[e.g.][]{2024A&A...690A.272K}.

While that is certainly a viable solution, recent data indicate that maybe not all disks around massive stars are that large and/or massive (e.g., \citealt{ginsburg2023}). Furthermore,  simulations that do not even resolve the disks find similar higher multiplicity for high-mass regions as observations do (e.g., \citealt{guszejnov2023}). However, as the authors point out, if they correct for observational incompleteness, the predicted multiplicity is below the observational constraints, especially in the high-mass range, likely because of not resolving the disks and hence no disk fragmentation in the simulations. Therefore, disk fragmentation appears to be an important ingredient to explain the higher multiplicity of high-mass stars, but it may not be the only solution, and other formation paths need further investigations as well.

In the following, we will concentrate on the larger-scale fragmentation of the parental gas clumps, relating to the filament and clump/core fragmentation. To characterize these fragmentation properties, one needs to differentiate various physical processes contributing to the fragmentation properties, in particular, thermal Jeans fragmentation, additional contributions from the turbulent motions, the magnetic field or the initial gas density distributions (see also sections \ref{density} and \ref{magnetic}). 

In the classical Jeans formulation, thermal clouds have critical mass and length scales above which they are not stable anymore but fragment (e.g., \citealt{jeans1902,stahler2005}). Focusing on the Jeans-length $\lambda_J$, this is proportional to the sound speed (that depends on the square root of the temperature $T$) and inversely proportional to the square root of the density $\rho$. Hence, $\lambda_J\propto \sqrt{T/\rho}$.

\begin{figure}[ht]
\includegraphics[width=0.99\textwidth]{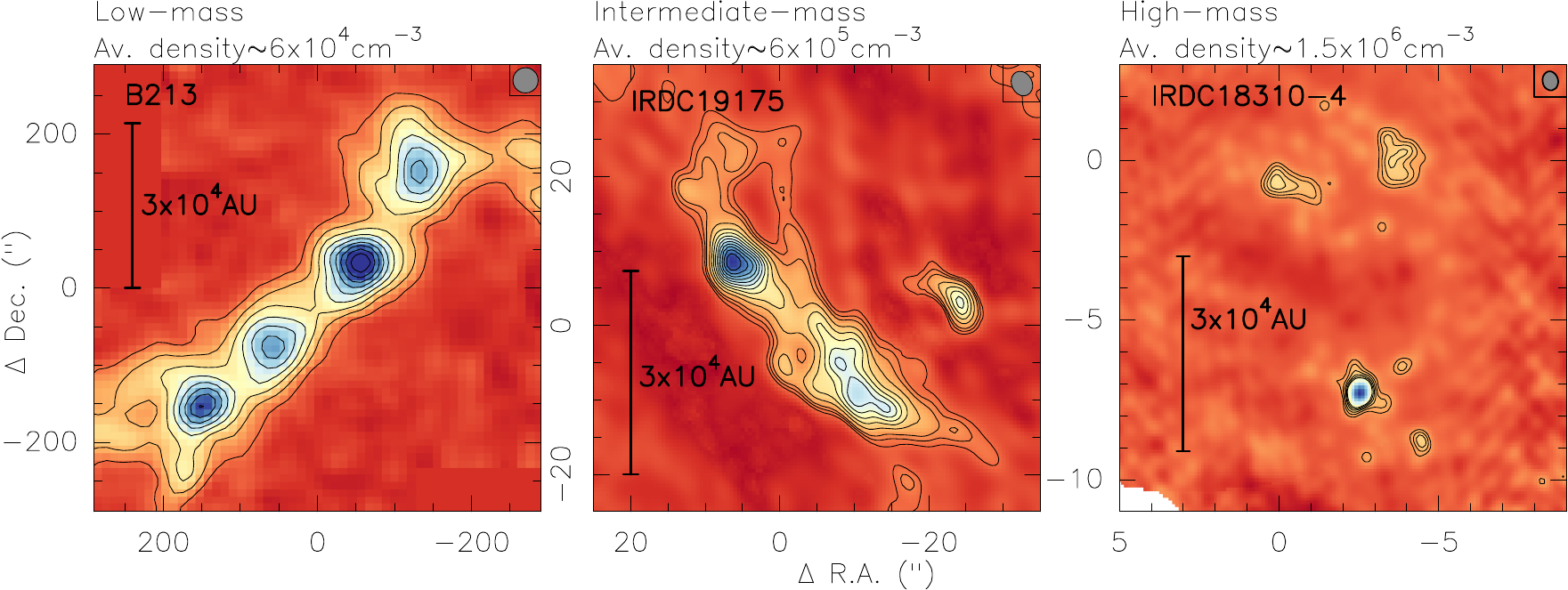}
\caption{Comparison of core separations and average densities for example low-, intermediate and high-mass star-forming regions on the same physical scales. The left, middle and right panels show data in color and contours for B213 (N$_2$H$^+$(1--0), \citealt{tafalla2015}), IRDC\,19175 (N$_2$H$^+$(1--0), \citealt{beuther2009}) and IRDC\,18310-4 (1.1\,mm continuum, \citealt{beuther2015}). Linear scale-bars and spatial resolution elements are shown to the left and top-right in each panel, respectively.}
\label{jeans-length}
\end{figure}

Setting this Jeans length into context with typical conditions in star-forming regions, one can estimate $\lambda_J$ for typical densities and temperatures in low- and high-mass star-forming regions. Following the original comparison in \citet{beuther2013}, Figure \ref{jeans-length} shows fragmentation data for very young example star-forming regions at low- (B213 in Taurus), intermediate- (IRDC\,19175) and high-mass (IRDC\,18310-4) at comparable scales and linear resolution elements \citep{tafalla2015,beuther2009,beuther2015}. While all three regimes form multiple, cluster-like regions, the core separations vary significantly among them. Typical separations for the low- and intermediate-mass regions are $\sim$22000\,au and $\sim$5000\,au, respectively. The high-mass region shows different hierarchical levels of fragmentation with the smallest separations almost at the spatial-resolution limit of $\sim$2600\,au. For estimates of the Jeans-length, one needs estimates for the temperatures and densities of the parental gas structures in the regions. Assuming a temperature of 15\,K at the onset of star formation, and using the reported average densities of the corresponding parental structures as presented in Fig.~\ref{jeans-length}, the respective Jeans-lengths for the low-, intermediate- and high-mass regions are $\sim$20000, $\sim$6000 and $\sim$4000\,au. While the overall size of the regions is rather similar (50000\,au or 0.25\,pc), the core separations vary tremendously. Nevertheless, the fragment separations of all mass regimes roughly correspond to the scales one obtains from the comparatively simple Jeans analysis. For more dicussion also about filament fragmentation, we refer to \citet{hacar2023} and \citet{pineda2023}.

While the above analysis just selected individual regions of different mass, sample studies also investigated the fragmentation of cluster-forming gas clumps. Some early studies targeting infrared dark clouds found slightly larger core separations that agreed better with turbulent Jeans fragmentation (e.g., \citealt{pillai2011,wang2014}). In this scenario, the line-width for estimating the Jeans-length is not the thermal line-width but that produced by turbulent motions within the regions (see also section \ref{turbulence}). Similarly, the study of different evolutionary stages by \citet{avison2023} is also more consistent with turbulent Jeans fragmentation. In contrast, some other sample studies targeting either earlier (e.g., high-mass starless core candidates) or later evolutionary stages (e.g., high-mass protostellar objects) find their regions again to agree more with thermal Jeans fragmentation (e.g., \citealt{palau2013,palau2014,palau2015,palau2018,beuther2018,sanhueza2019,svoboda2019,morii2024}). Recently, \citet{traficante2023} show that towards different evolutionary stages the average core separations decrease with time. This may be interpreted in a way that the whole star-forming regions continue to globally collapse during their evolution. The latter picture appears plausible and is able to explain many of the observations. However, the fact that independent of evolutionary phases, some studies find separations consistent with thermal Jeans fragmentation and other favour a turbulent contribution, indicates that the fragmentation processes are not universally always dominated by the same processes, but that the contributions from gravity and turbulence may vary from region to region, also depending on the environment (see also Sect.~\ref{turbulence}).

In addition to gravity and turbulence, the magnetic field may influence the fragmentation of star-forming regions (see also Sect.~\ref{magnetic}). Simulations indicated that regions with higher magnetic field strength may fragment less (e.g., \citealt{commercon2011,commercon2022}). This process has now been started to be investigated by interferometer polarization studies of samples of high-mass star-forming regions, and early results are not conclusive yet. While \citet{zhang2014} or \citet{palau2021} found a tentative correlation between the number of fragments and the mass-to-magnetic-flux ratio $(M/\Phi_B)$, the study by \citet{beuther2024} could not confirm that yet. So far, high-spatial-resolution magnetic field studies are still only scratching the surface (see also section \ref{magnetic}), and many more results are expected in the coming years.

Altogether, these data indicate that the initial fragmentation of star-forming regions may be influenced by various parameters, in particular gravity, turbulence and magnetic fields. While average densities and by that Jeans length vary between low- and high-mass star formation, we do not find qualitatively significant differences between the fragmentation properties from low- to high-mass star-forming regions. 

\subsection{Disk formation and disk properties}
\label{disks}

Disks around protostars are a natural byproduct of star formation from low-mass to high-mass 
(e.g., \citealt{2013A&A...560A.103M, zotero-2040, beltran2016, 2017ApJ...834..178Y, cesaroni2017, 2022ApJ...929...76S, ahmadi2023,tobin2024}).
The infalling gas from large scales will speed up its rotation due to the conservation of the net angular momentum from the large-scale inflow until gravito-centrifugal equilibrium is reached at disk radii; the actual equilibrium can contain further contributions such as thermal pressure, magnetic pressure, turbulent pressure, radiation pressure, or ram pressure.

Hence, a variety of phenomena are common for both, low- and high-mass disk formation and evolution:
disks are initially growing in size due to angular momentum inflow from larger scales, 
their further evolution is regulated by fluid instabilities and MHD disk winds giving rise to angular momentum transport, 
accretion from the envelope onto the disk and accretion from the disk onto the star is in general variable in nature, 
even accretion bursts have been observed in both regimes, and
especially if born in a cluster, the environment will impact its long-term evolution due to encounters and/or feedback.
Besides these basic similarities, the details can be expected to vary from low- to high-mass disks, such as the main driver(s) of accretion and the cause of accretion bursts.

Circumstellar disks cannot grow in mass freely in general. 
Even in a scenario where the mass infall rate onto the disk is higher than the current accretion rate through the disk and/or onto the protostar, the increase in disk mass, or more specifically the higher disk-to-star mass ratio (a proxy for the global Toomre parameter of the disk, \citealt{1964ApJ...139.1217T}) makes it more prone to gravitational instability, yielding the formation of spiral arms, which enhance the disk accretion rate by acting gravitational torques.
That means that self-gravitating disks are entering such a self-regulated mode under those circumstances.
This has been demonstrated in theoretical work by \citet{2004MNRAS.351..630L} and \citet{zotero-1446} for the low-mass regime; see also the review by \citet{kratter2016} and \citet{kuiper2011} for the high-mass regime.

The detection of rather large disks or disk candidates and rotating tori on larger scales \citep{beltran2016} supports that disk radii have a tendency to increase with stellar host mass.
Since the review of \citet{beltran2016}, a couple of disks have been detected surrounding high-mass protostars, all with sizes way larger than in the low-mass regime; see below for a discussion on disk substructure.
A counterexample might be the ALMA observation (continuum and molecular emission) of the W51 star-forming complex by \citet{2020ApJ...905...25G}, although a circumstellar disk of a maximum size of $500-100 \mbox{ au}$ cannot be ruled out yet, and the existence of a collimated outflow points toward disk accretion (see also \citealt{ginsburg2023} for intermediate disk sizes).

Although the formation and evolution of disks is controlled by angular momentum transport in all ranges of stellar masses, the underlying physical processes might differ in the low- and high-mass regime. 
A common process is the removal of the disk angular momentum through the MHD disk winds, as demonstrated by early simulation work in the low-mass regime \citep[e.g.][]{shibata1985, 1986PASJ...38..631S, 1994ApJ...433..746S, 2022MNRAS.512.2290T} and recent studies in the high-mass regime \citep[e.g.][]{oliva2023a, oliva2023b}.
The latter simulation studies were used for a quantitative comparison with observational data in \citet{2023A&A...680A.107M}, because the expected gas velocities along the streamlines could be detected in \citet{2022NatAs...6.1068M} thanks to below sub-au resolution in Very Long Baseline Interferometry (VLBI) observation of water maser emission.

Further mechanisms of angular momentum transport, especially in evolved stages, might be different in low- and high-mass disk evolution. While accretion through massive disks can be efficiently driven by gravitational torques \citep[e.g.][]{zotero-988, kuiper2011}, a variety of different instabilities have been proposed for angular momentum transport in low-mass disks:
convective instability \citep{zotero-2173, zotero-2255},
baroclinic instability \citep{zotero-966},
magneto-rotational instability \citep{zotero-2515, zotero-1920}, and 
self-gravitating instabilities \citep{zotero-1406}. 
%self-gravitating instabilities \citep[e.g.][]{zotero-1406, 2009ApJ...695L..53B}. 
The latter effect is only dominant for a high disk-to-star mass ratio, that is, more important during early formation times \citep[e.g.][]{2017ApJ...835L..11T, zotero-9172}.

The dominance of self-gravity in disks surrounding high-mass protostars is also responsible for another important difference in high-mass star formation compared to the low-mass case, namely the ubiquity of disk fragmentation resulting in the formation of close-separation multiples (see also Sects.~\ref{bimodal} and \ref{fragmentation}).
Although early numerical studies on cloud collapse and disk formation in the high-mass regime could not properly resolve the length scales of the physical processes for disk fragmentation, namely the Jeans length of the spiral arms and fragments that form within the spirals, this obstacle was overcome and its implications were studied in a thorough convergence check by \citet{2020A&A...644A..41O}.
These early results were supported by a comparison of two different grid techniques in \citet{2023A&A...672A..88M} and augmented by the effect of magnetic fields in
\citet{2021A&A...652A..69M, 2023A&A...673A.134M} and \citet{2022A&A...658A..52Ca}.
Hence, disk fragmentation seems to be a phenomenon that happens everywhere in high-mass star formation for different environmental conditions (e.g., \citealt{2022MNRAS.517.4795M}) and could naturally explain the high fraction of spectroscopic binaries in the high-mass regime \citep{2018MNRAS.473.3615M, 2020A&A...644A..41O}.
Similarly, in simulations of low-mass star formation by \citet{2019MNRAS.484.2341B}, disk fragmentation contributes significantly to the formation of small-separation binaries, but happens for a lower fraction of star-disk systems than in models of high-mass star formation.
This trend of ubiquitous spiral arm formation and disk fragmentation for high disk-to-star mass ratios is observationally confirmed, as most observations of resolved disks show spiral arm like substructure and/or embedded fragments \citep{zotero-3577, 2018ApJ...869L..24I, zotero-666, zotero-1704, zotero-2380, 2021A&A...655A..72S, 2021A&A...655A..84S, 2022NatAs...6..837L, 2023NatAs...7..557B}.

A phenomenon that is at least sometimes related to disk fragmentation and migration is the occurrence of strong accretion bursts during high-mass star formation \citep{2015MNRAS.446.4088T, 2016ATel.8732....1S, zotero-2120, zotero-1814, 2018ApJ...854..170H, 2018ApJ...863L..12L, burns2020,zotero-2959, 2021ApJ...922...90C, 2024arXiv240510427W}.
The analogous phenomena in the formation of low-mass stars, often referred to as FU Orionis variables (FUors) and EXor bursts, have been known for decades \citep{1977ApJ...217..693H, 1996ARA&A..34..207H, 2014prpl.conf..387A, 2019ApJ...872..183F, 2021ApJ...920..119L, 2022ApJ...937....6J, 2023ASPC..534..355Fa, 2024AJ....168..122P}.
The physical origin of bursts could not yet be determined and theoretical ideas cover a range of phenomena such as
gravitational instability \citep[e.g.][]{zotero-1670, 2011ApJ...729...42M, zotero-3041, 2020A&A...644A..41O, 2023MNRAS.518..791E},
gravitational instability plus magneto-rotational instability \citep[e.g.][]{2001MNRAS.324..705A, 2009ApJ...701..620Z, zotero-1955},
binarity \citep[e.g.][]{1992ApJ...401L..31B, 1996MNRAS.278L..23C, zotero-9142, 2020A&A...641A..59K},
tidal interactions with cluster stars \citep[e.g.][]{zotero-3210, 2008A&A...492..735P, 2020MNRAS.491..504C, zotero-1955, zotero-2138}, 
planet-disk interaction \citep[e.g.][]{2012MNRAS.426...70N}, thermal instability \citep[e.g.][]{1983MNRAS.205..359F, 1983MNRAS.205..487P, 1985MNRAS.212..105L, zotero-1162, zotero-810, 2021A&A...651L...3E} or thermal instability impacted by planets \citep[e.g.][]{2004MNRAS.353..841L}.

Based on the variety of even strong burst-triggering phenomena and the variety of accretion-driving angular momentum transport mechanisms outlined above, disk accretion can be expected to be at least moderately variable for most of the time and for the full range of stellar host masses.
For the so-called luminosity problem in low-mass star formation, i.e., the observed accretion luminosity of a star-forming region such as Taurus-Auriga \citep{1990AJ.....99..869K, 1995ApJS..101..117K} or Perseus, Serpens, and Ophiuchus \citep{zotero-3174} or individual clouds in Perseus, Serpens, Ophiuchus, Lupus, and Chamaeleon \citep{evans2009} is lower than expected when assuming a constant disk-to-star accretion rate during the embedded phase \citep{1990AJ.....99..869K, zotero-3174}, variability in the accretion history is an obvious solution, because proto-stars are expected to spend most of their time in a mode of low accretion rates.
This interpretation is also in line with observed individual objects which are ''under-luminous`` with respect to their other evolutionary indicators such as their outflows \citep[e.g.,][]{2006ApJ...651..945D}, and the substructure of outflows indicate variable ejection themselves.

Disk lifetimes can be expected to decrease significantly with increasing stellar host mass.
First, the dynamical timescales of disk evolution decrease with stellar mass,
secondly, as discussed above, the main accretion driving mechanisms seem to be more efficient for high-mass disks compared to low-mass disks, and
thirdly, the impact of feedback by the growing host star and the cluster environment will become stronger for high-mass sources (e.g., \citealt{hollenbach1994,winter2020}). 
A recent study by \citet{vanterwisga2023} shows that the disk masses depend on the radiation field and decrease with increasing irradiation. However, it remains to be seen whether the disk masses or sizes themselves are critical in star and planet formation processes since recent JWST observations have shown that the inner disk properties (important for rocky planet formation) in extreme UV environments like NGC6357 appear similar to those in nearby relatively isolated low-mass regions \citep{ramirez-tannus2023}.

\subsection{Jet and outflow properties}
\label{outflows}

From a theoretical point of view, the physics of jets and outflows is best discussed by starting from the smallest launching scales and going up to the large-scale phenomena. On the smallest scales, we can expect the launching, acceleration, and collimation of a fast jet due to the mechanisms outlined in \citet{zotero-3111}, accommodated by magnetic pressure-driven disk winds or tower flows \citep{zotero-2692}; slow disk winds and tower flows are also predicted via the \citet{zotero-3111} mechanism by \citet{zotero-2266}.
Since the gas density is a scale-free parameter in the associated ideal MHD equations, the only further difference from low- to high-mass star formation is given by the stellar mass itself.
A higher mass of the protostar yields a higher gravitational potential, a higher escape speed, a higher value of Keplerian rotation at the (identical) launching radius, and hence, eventually a higher jet velocity \citep{zotero-2266}.
So, we would expect an increase of accretion rate, outflow rate, and jet velocity toward higher-mass stars.
This idealized picture will be modified by including effects of non-ideal, resistive MHD in combination with cosmic ray ionization and photoionization as well as differences in disk thermodynamics and disk evolution.

\begin{wrapfigure}{r}{0.6\textwidth}
\includegraphics[width=0.59\textwidth]{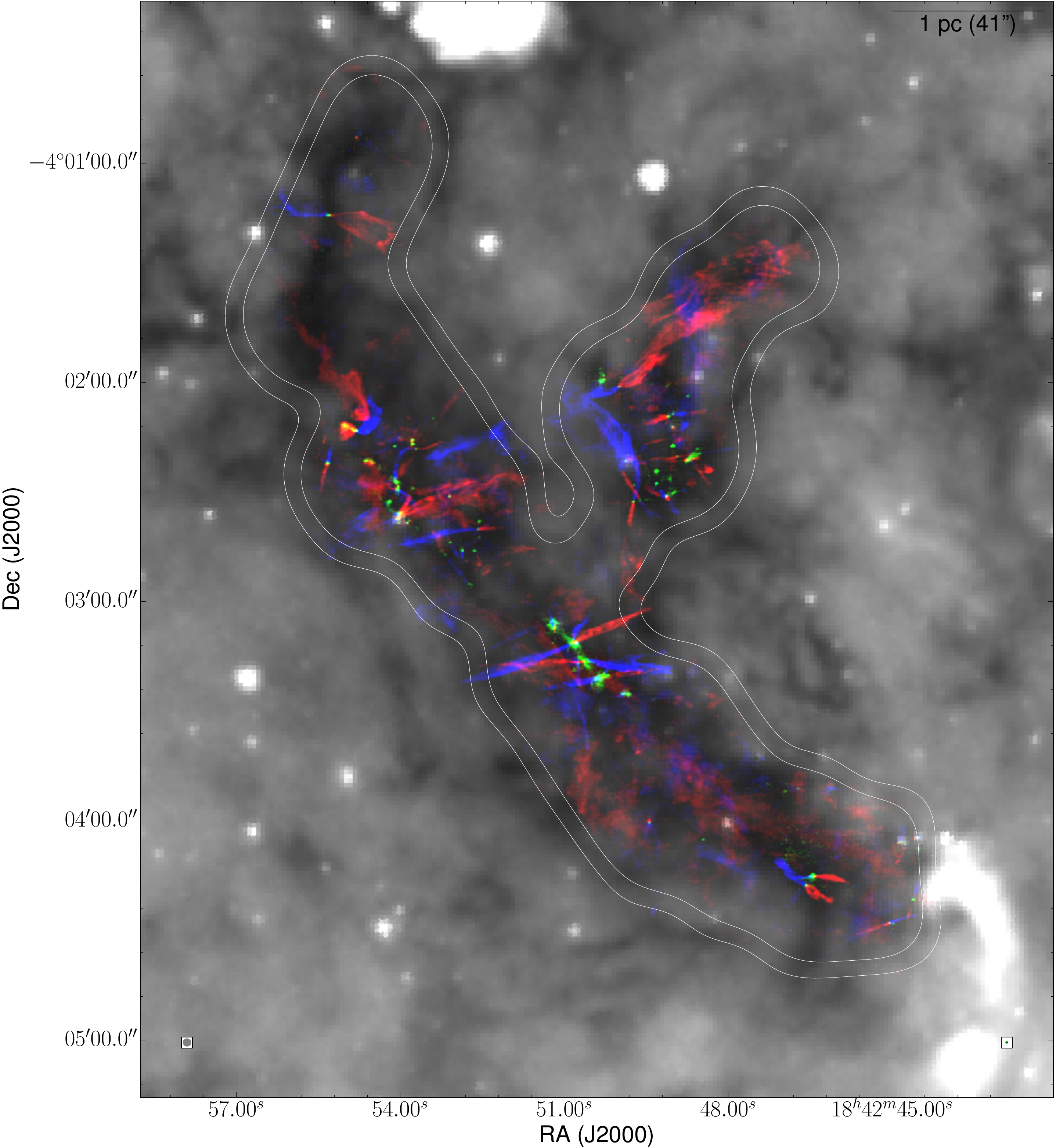}
\caption{Example for the widespread outflow distribution in an infrared dark cloud from \citet{kong2019}. While the grey-scale shows the Spitzer 8\,$\mu$m emission, the blue and red structures present the blue- and red-shifted CO(2--1) outflow emission. The white contours outline the ALMA primary beam response at the 30 and 50\% level. While a scale bar is shown in the top-right, the Spitzer and ALMA spatial resolution elements are shown at the bottom-left and right, respectively.}
\label{g28_outflows}
\end{wrapfigure}

MHD-driven outflows are a ubiquitous phenomenon in simulations of
low-mass \citep[e.g.][]{zotero-2879, zotero-2955, zotero-1017, zotero-1729, zotero-3485, 2016A&A...587A..32M} and
high-mass star formation \citep{zotero-1921, zotero-2650, zotero-2512, zotero-261, zotero-34, zotero-1122, zotero-3440, matsushita2017, zotero-2178, zotero-3097, commercon2022, oliva2023b}, although in the latter case high-velocity jets ($> 100 \mbox{ km}/\mbox{s}$) are only observed in simulations starting from a disk configuration \citep{zotero-261} or in simulations of the highest resolution core collapse \citep{matsushita2017, zotero-2178, oliva2023b}.

On larger scales, MHD-driven jets and disk winds impact the proto-stellar surrounding, but are also impacted in their propagation and collimation by the infalling environment. Here, high-mass star formation may differ due to higher ram pressure of the infalling gas of higher density, but studies suggest that the entrainment efficiency measured as the effective ejection-to-accretion rate \citep{zotero-34, matsushita2017, zotero-2178, oliva2023b} is not different from the low-mass case \citep{tomisaka2002, zotero-2955, 2020ApJ...900...15L}. 
As expected, in all cases, the outflow opening angles widen with time, although the main driving mechanism behind the increase in opening angle depends on the physics included in the simulations such as continuum radiative forces, line-driven forces, photoionization, and the strength of infall.
Later in their evolution, the bipolar regions around high-mass protostars differ from their low-mass counterpart due to the early growth of H{\sc ii} regions \citep{kuiper2018}, as described in more detail in Sect.~\ref{feedback}.

The study of the orientation of multiple outflows in a filament, in a filament-hub system, and/or during cluster formation might tell us something about the ratios of gravity, turbulence, and magnetic fields in these structures (e.g., Fig.~\ref{g28_outflows}). Unfortunately, no numerical simulation on these scales was able to self-consistently resolve the MHD-launching of jets. In order to study their feedback, modelers rely on injecting them based on a sub-grid model, even in the case of MHD simulations. 

From an observational point of view, a scaling of higher outflow parameters towards higher-mass star-forming regions has been detected. Early studies of low-mass regions (e.g., \citealt{cabrit1992,bontemps1996,bachiller1999}) were soon complemented by investigations of high-mass regions (e.g., \citealt{shepherd1996,henning2000,beuther2002b}), and a first outflow parameter investigation summarizing all outflow studies at that time was presented in \citet{wu2004} (see also more recent studies by \citealt{lopez2009,maud2015}). These studies consistently show that parameters like the outflow mass, force and also the rate of entrained gas all scale with the luminosity of the star-forming region. In particular, the outflow rate is important because it is directly linked to the accretion rate (e.g., \citealt{beuther2002b}). Therefore, the observed increase of outflow rate with protostellar luminosity indirectly shows that also the accretion rates should increase with increasing protostellar mass (e.g., \citealt{richer2000,arce2007,frank2014}).

Another question is whether the collimation of high-mass outflows may be lower than that of their low-mass counterparts (e.g., \citealt{shepherd1998,ridge2001,wu2004}). Initial studies of high-mass regions lacked a consistent comparison of evolutionary stages, also taking into account the much larger distances of high-mass regions. To overcome such inconsistencies, \citet{beuther2005} compiled the existing high-resolution studies and found that collimation factors at the earliest evolutionary stages were the same for low- and high-mass outflows.  Only at later evolutionary stages, the high-mass outflows become less collimated. Therefore, they advocated that high-mass protostars have also collimated MHD-driven jet-like outflows at the early evolutionary stages, and only later, when winds and ionizing radiation become important, high-mass outflows may become less collimated (see also \citealt{arce2007}).

In summary, outflow parameters clearly scale with protostellar luminosity, which is indicative of increasing accretion rates with mass. Nevertheless, the underlying physical processes, especially at the early evolutionary stages, namely MHD-driven disk winds appear to be the same for low- and high-mass protostars. Only at later evolutionary stages, when the high-mass protostars form winds and ionizing radiation, the characteristics of the outflows start to change.

\subsection{Feedback}
\label{feedback}

As mentioned earlier, the effect of feedback in its impact on shaping the IMF is discussed in the context of the environment (Sect.~\ref{bimodal}) and not repeated here. We just note that recent modeling progress has been achieved on this aspect in \citet{2022MNRAS.512..216G, 2022MNRAS.515.4929G, 2023OJAp....6E..48G, 2024MNRAS.527.6732F, 2024ApJ...973...40K}, and we refer the reader to the recent review by \citet{2024ARA&A..62...63H}.

Feedback mechanisms common to low- and high-mass stars are radiative heating and jets/outflows.
Feedback effects unique to high-mass stars are continuum radiation forces, photoionization feedback and H{\sc ii} regions, UV-line-driven forces and stellar winds as well as supernovae.
Supernovae are certainly a strong feedback on the galactic evolution of the interstellar medium, and hence, affect the environmental conditions for the next generation of stars, but will not be further discussed here in the context of a comparative view of the star formation process itself.
Stellar winds are commonly also exclusively treated as the last pre-supernovae feedback effects, which do affect the star formation process itself only to a minor degree (e.g., \citealt{2022MNRAS.515.4929G}). 
But the underlying UV-line-driven forces greatly impact at least the final stages of disk accretion for high-mass stars and represent an intrinsic upper mass limit for stellar growth via disk accretion \citep{2019MNRAS.483.4893K}. The value of this limit depends on the accretion rate \citep[][their Fig.~4]{2019MNRAS.483.4893K} and metallicity \citep[][their Fig.~6]{2019MNRAS.483.4893K}. Observationally, a first candidate of such a force-driven disk ablation scenario has been detected as an SiO layer in \citet{zotero-872}.
On scales beyond disk sizes, an early launching of stellar winds can quench the accretion flow onto the star-disk system \citep{rosen2022}.

Besides stellar wind driving, radiation impacts the stellar environment through photoionization and radiative forces.
The ionizing feedback from high-mass star-forming regions is a significant difference between the low- and high-mass star formation regime. In the early works about ultracompact H{\sc ii} (UCH{\sc ii}) regions, it was argued that the formation of such UCH{\sc ii} regions may terminate the accretion process (e.g., \citealt{kurtz1994,churchwell2002}). However, soon after hypercompact H{\sc ii} (HCH{\sc ii}) regions were identified, differing significantly from the previous UCH{\sc ii}s. In particular, the HCH{\sc ii} regions are $\geq$10 times smaller ($\leq 0.01$\,pc and $\sim$100 times denser than the UCH{\sc ii} regions (e.g., \citealt{kurtz2002,beuther2007,hoare2007}). Important for the formation of high-mass stars is that these HCH{\sc ii} regions may be gravitationally trapped or quenched, and that the accretion processes can continue through the HCH{\sc ii} regions as ionized accretion flows (e.g., \citealt{walmsley1995,keto2002,keto2003}). Similar to the gravitational trapping, the accretion can continue further through accretion disks also in the ionized phase (e.g., \citealt{sollins2005,keto2006,keto2007,kuiper2018,galvan-madrid2023}). While this ionizing radiation and trapping of the HCH{\sc ii} region is a clear difference between the low- and high-mass regime, it is interesting to point out that the actual accretion processes through disk-like structures remain qualitatively similar. The ionizing radiation from the central forming massive protostar likely escapes the region preferentially through already carved out cones from the bipolar jets and outflows (see, e.g., \citealt{tan2003b,tan2014,zotero-1344,zotero-314, kuiper2018, 2019ARA&A..57..227K}). Furthermore, the ongoing accretion onto the H{\sc ii} region can change the local density and, because recombination scales quadratic with density, may cause a "flickering" and changes of the ionized emission on comparably short timescales (e.g., \citealt{franco-hernandez2004,galvan-madrid2008,peters2010,depree2014,depree2018}).

A different feedback process from larger evolving H{\sc ii} regions can be constructive as well as destructive for future star formation processes. Expanding H{\sc ii} regions form large bubble-like structures that can destroy environmental clouds. However, also the other way round, expanding H{\sc ii} regions may push gas together and even foster new star formation processes (e.g., \citealt{kendrew2016,palmeirim2017,luisi2021}). 
However, since these new regions show no significantly different star formation signatures compared to other star-forming regions, triggering  does not appear to be a dominant star-formation effect but one of the many processes that contribute to it (see \citealt{elmegreen2011} for a full review).

As discussed in \citet{megeath2022}, the star formation efficiency per free-fall time barely varies between clouds forming low- or high-mass stars, indicating that it does not depend strongly on the feedback processes from the high-mass stars.

One distinct difference between the pre-ZAMS evolution of low-mass and high-mass stars is that the accretion rate $\dot{M}$ during high-mass star formation can be so high that the accretion timescale $t_\mathrm{acc} = M_\star / \dot{M}$ becomes shorter than the Kelvin-Helmholtz contraction timescale $t_\mathrm{KH}$ (when evaluating the ratio of these two timescales, one should keep in mind that both timescales are themselves time-dependent quantities of the evolving protostar, which can change by several orders of magnitude rapidly, e.g., \citealt{zotero-2172}).
As a result, the protostar bloats up its atmosphere significantly, reaching stellar radii of order $100 \,\mbox{R}_\odot$.
This mechanism holds for spherical accretion \citep{zotero-2172}, disk accretion \citep{zotero-2196}, and self-consistent accretion during cloud collapse and disk formation \citep{zotero-3598, zotero-809}.
The condition $t_\mathrm{acc} < t_\mathrm{KH}$ can even be fulfilled after the high-mass star has contracted down to the ZAMS, e.g., during an accretion burst as a result of disk fragmentation (see also Sect.~\ref{disks}).

The bloating of the stellar atmosphere changes the stellar radius and photospheric temperature, its bolometric luminosity can be expected to remain unaffected, but this might as well depend on its internal structure (e.g., \citealt{linz2009,zotero-809}).
Importantly, during the bloated stage, the spectrum of the (proto)star shifts toward the infrared \citep[e.g.][their Fig.~11]{zotero-3598}.
As a consequence, photoionization feedback and the formation of an H{\sc ii} region will be delayed or suppressed.
In the case of the formation of the first high-mass stars in the universe, where photoionization feedback dominates over continuum radiation forces, the bloating effect and suppressed photoionization can be crucial to determine the intrinsic stellar upper mass limit \citep{zotero-2861}.

Additionally, continuum radiation forces effectively lower the gravitational attraction of a forming star, a feedback effect unique to high-mass stars.
As a zero order estimate, the direct radiative force of a forming star becomes comparable to its own gravitational attraction, and hence could be able to redirect an accretion flow into an outflow, when the protostar has reached about $20 \mbox{ M}_\odot$ \citep{zotero-1845, zinnecker2007}. 
Due to the fact, that the growing radiation force first has to slow down the infalling environment around the forming star before being able to launch an outflow, the proto-star can still reach about $40 \mbox{ M}_\odot$ as its maximum mass \citep{kahn1974, zotero-2625}. 
\citet{zotero-2066} put forward the idea that classical disk accretion, which low-mass and high-mass stars seem to have in common, might solve the so-called radiation pressure problem for the formation of the more massive stars due to the strong anisotropy of the radiation field introduced by the high optical depth of the disk. 
In radiation transport analyses of a star-disk-outflow configuration, \citet{krumholz2005b} confirmed such an anisotropy of the resulting stellar radiation field.
In direct gravito-radiation-hydrodynamical experiments of this scenario, 
\citet{yorke2002} followed the star and disk formation process up to $43 \mbox{ M}_\odot$ when including frequency-dependent radiation transport, and eventually
\citet{kuiper2010} demonstrated that this effect indeed circumvents the radiation pressure problem at least up to $150 \mbox{ M}_\odot$. 
Numerically this requires a high spatial resolution of the inner disk region, where the stellar radiation encounters the high optical depth of the accretion disk \citep{kuiper2010}. 
Hence, although continuum radiation forces denote a feedback effect onto the nearby environment of forming high-mass stars, they do not significantly alter the qualitative scenario of the formation of stars in general.

\subsection{Chemistry}
\label{chemistry}

The main constituents of molecular clouds (H$_2$ and He) are largely unobservable under typical gas conditions, so the study of the gas in star-forming regions needs to rely on alternative species whose abundance is not only low but potentially sensitive to changes in the physical state of the gas. As a result, interpreting gas observations requires understanding the diverse and often complex chemical processes that affect the abundance of molecules during star formation. In this section we briefly summarize the dominant processes that control the abundance of the main molecular tracers of the star-forming gas. A more detailed view of the chemistry of star-forming regions can be found in the rich literature available (e.g., \citealt{bergin2007,herbst2009,caselli2012, ceccarelli2014,jorgensen2020}).

In broad terms, two distinct chemical phases can be distinguished during the process of star formation. Prior to the birth of a protostar, the gas conditions are characterized by a gradual increase in the density and a decrease of the temperature in the absence of heating sources. After star formation begins, the chemistry of the surrounding gas is modified by the feedback from the protostar, which includes direct heating, shock-acceleration from outflows and winds, and photoprocessing from the stellar radiation. We briefly review the main characteristics of these two phases.

Pre-stellar conditions are characterized by a significant increase in the density of the gas with respect to its surroundings. In regions of low-mass star formation, this increase approximately coincides with the gas reaching a critical density for the freeze out of most molecular species onto the cold dust grains. As a result, low-mass starless cores develop a stratified internal chemical composition where the interior becomes depleted of CO and other C-bearing species, while N-bearing species such as N$_2$H$^+$ and NH$_3$ remain in the gas phase with little or no abundance change \citep{bergin2007,aikawa2008,caselli2012}. 
In addition, due to the lower zero-point energy of deuterium-substituted species, the abundance of deuterated molecules in the pre-stellar gas can increase several orders of magnitude over the D/H cosmic ratio \citep{crapsi2005,ceccarelli2014}.  Similar chemical changes seem to characterize the pre-stellar phase of high-mass star formation, although the study of this phase has been less extensive than for low-mass regions (e.g., \citealt{pillai2006,fontani2008,gerner2014,gerner2015,gieser2022,gieser2023}). IRDCs with no evidence of star formation are the most likely counterparts of the low-mass pre-stellar phase, and frequently show evidence for CO freeze out and deuterium fractionation. Due to the higher gas density of these clouds compared to low-mass star-forming regions, the effect of freeze out and deuteration dominates the composition of regions larger than the individual cores, and frequently extends to scales of clumps or even a significant fraction of the whole cloud (e.g., \citealt{hernandez2011,fontani2012,barnes2016}).
Complementary observations of the ice component also point to a similar composition of the prestellar material in regions of low- and high-mass star formation 
\citep{oberg2011,boogert2015}. 
Further progress in the study of this component is expected over 
the next few years thanks
to the increase in sensitivity brought by the JWST, which is rapidly expanding the inventory of molecular species detected in ice form \citep{mcclure2023,rocha2024}.

The formation of a protostar changes significantly the chemical composition of the surrounding gas. For low-mass protostars, the effect is limited by the low energy output of the sources, but it can be readily detected toward some of the youngest systems in the form of hot corinos, which are regions of high abundance in saturated organic molecules such as methanol, methyl formate, and dimethyl ether \citep{herbst2009,caselli2012, jorgensen2020}.
These hot corinos have typical sizes of 100\,au, and their rich chemistry is believed to be driven by the heating and evaporation of ice mantles from the dust grains at temperatures in excess of 100\,K (e.g., \citealt{ceccarelli2023}).
Well studied hot corinos include those associated to the double protostars IRAS 16293–2422 
\citep{cazaux2003,jorgensen2016} and NGC 1333 IRAS4A \citep{bottinelli2004,lopezsepulcre2017}, and sensitive surveys using ALMA are rapidly increasing the number of objects with hot corino chemistry \citep{yang2021,hsu2022}.
At larger distances from the protostar, shocks from outflows often enhance the abundance of species such as SiO, CH$_3$OH, and H$_2$O, especially in the class 0 phase (e.g., \citealt{bachiller1997,tafalla2010,kristensen2012,vandishoeck2021}).

As expected from their higher energy output, high-mass protostars stars have a stronger chemical effect on the environment. 
Their associated hot cores are significantly larger ($\sim10^4$\,au) and  typically warmer ($\sim300$\,K) than the low-mass hot corinos (e.g., \citealt{herbst2009}), but when the molecular ratios of hot cores and hot corinos are compared, nearly constant values are found over multiple orders of magnitude in source luminosity, pointing to a common chemistry that may be related to a similar pre-stellar phase \citep{coletta2020,nazari2022}. In contrast to the hot corinos, however, the hot core phase around a high-mass protostar is followed by the development of an UCH{\sc ii} region driven by the UV radiation from the central object. This development marks the beginning of the end for the molecular phase surrounding a high-mass protostar since the high overpressure of the ionized gas and the continuous stellar radiation leads to the inevitable expansion of the nascent H{\sc ii} region.
Apart from this later evolution, the chemistry of low- and high-mass star-forming regions seems remarkably similar, with most differences between the regions seeming to simply arise from their difference in density and temperature.

\begin{table}[ht]
\tabcolsep7.5pt
\caption{Summary of properties}
\label{summary_tab}
\begin{center}
\begin{tabular}{l|c}
\hline
Parameter & Properties\\
\hline
            &  Similarities from low- (lm) to high-mass (hm) star formation \\
\hline
Environment & lm: distributed \& clustered $|$ hm: clustered only \\
& IMF\&CF-lm(distributed) $\approx$ IMF\&CF-hm(clustered)\\
          
Density slope & lm $\approx$ hm \\
Turbulent/thermal support & lm $\approx$ hm\\
Variability & lm $\approx$ hm \\
\hline
& Quantitative differences\\
\hline
Infall/accretion &  $\dot{M}$-lm $<\dot{M}$-hm \\
Column density threshold & $N$-lm $<N$-hm  \\
Volume density threshold & $\rho$-lm $<\rho$-hm  \\
Magnetic field & cloud-scale: $B$-lm $\leq B$-hm \\  
               & core-scale: $(M/\Phi_B)$-lm $\sim 1 \leq (M/\Phi_B)$-hm\\
Disks & $R_\mathrm{disk}\mbox{-lm} < R_\mathrm{disk}\mbox{-hm}$ \\
Outflows & $\dot{M}_{\rm out}$-lm $<\dot{M}_{\rm out}$-hm\\
Fragmentation & $\lambda_J(\rm{lm})>\lambda_J(\rm{hm})$\\
Chemistry & freeze out spatial scale: core-lm $<$ clump/cloud-hm\\
        & sublimation spatial scale: hot corino-lm $<$ hot core-hm \\
Multiplicity & lm $<$ hm \\
\hline
& Qualitative differences \\
\hline
Feedback & lm \& hm: radiative heating and jets/outflows\\ 
 & hm: supernovae, stellar winds, radiation forces, photoionization\\
& $\rightarrow$ can be constructive and destructive\\
& disk-mediated accretion: lm (molecular); hm (also ionized)\\
\hline
\end{tabular}
\end{center}
\begin{tabnote}
Table notes: lm and hm refer to low-mass and high-mass star formation. CF: companion fraction. $(M/\Phi_B)$: mass-to-magnetic-flux ratio in units of the critical mass-to-flux ratio
\end{tabnote}
\end{table}

\section{Synthesizing a comparative view}
\label{sythesize}

While the past decades in star formation research often stressed a kind of bimodality between low- and high-mass star formation, we can now refine that picture using the different results discussed in section \ref{properties}. A short summary of these results is presented in Table \ref{summary_tab}. In general, one can separate three different categories: (a) close similarities, (b) quantitative differences and (c) qualitative differences.

\subsection{Similarities}

Let's start with the similarities between the low-mass and high-mass star formation regime. One of the early claims in high-mass star formation research was that massive stars form from turbulent initial conditions. As outlined in section \ref{turbulence}, when one looks at the earliest evolutionary stages in IRDCs with high enough spatial and spectral resolution, the observed line-widths are similarly narrow as their low-mass counterparts. Most observations find non-thermal velocity dispersions in the subsonic to transonic regime. Hence, the early assumption of high-mass star formation happening mainly in more turbulent clouds seems no longer valid. This is also supported by the characterization of the fragmentation of high-mass regions where many studies are consistent with thermal Jeans fragmentation (section \ref{fragmentation}).

While early theories predicted different density profiles for high-mass regions compared to their low-mass counterparts (section \ref{density}), observations over the past decades have shown that the observed density profiles $\rho\propto r^{-p}$ from low- to high-mass star-forming regions typically all have slopes with $p$ varying only between roughly $1.5$ and $2$, almost independent of mass. This indicates that the actual star formation process is comparably insensitive to the actual density structure. In that picture, the power-law slope of $\sim2$ may be an "attractor" where potentially initially flatter density distributions may typically approach $p\approx 2$ slopes (section \ref{density}).

One also wondered whether the environment plays an important role in the star formation processes. While there are certainly huge differences in many of the physical properties of the environment, e.g., external radiation fields, nearby H{\sc ii} regions or supernovae events, associated pressure difference, arm or interarm location, the outcome of the star formation process appears to not vary that much. As outlined in section \ref{environment}, critical characteristics of the final clusters, in particular the IMF and multiplicity appear to not vary significantly with the environment (as stated in Sect.~\ref{intro}, we do not review very extreme environments like the CMZ or galactic mergers here). 
Furthermore, while disk masses depend on the irradiation from nearby massive stars, it remains to be shown how important such processes are for planet formation since recent studies indicate inner disk properties in extreme UV environments to be similar to those in nearby low-mass regions (section \ref{disks}).

Variability is one of the less explored aspects. Variations in luminosity are observed both toward low- and high-mass protostars, suggesting that both regimes undergo variations in the rate of accretion (section \ref{disks}). So far, no obvious differences in the variability of low- and high-mass regions have been identified, although the statistics are still poor. Studies expected in the coming years will help clarify this issue.

\subsection{Quantitative differences}
\label{quantitative}

There are many physical and chemical properties within star-forming regions that vary quantitatively from low- to high-mass. Since the main accretion phases in low- and high-mass star formation typically last for similar time-scales of several $10^5$\,yrs (e.g., \citealt{mckee2002,evans2009}), one expects higher accretion rates in high-mass star-forming regions. 
Early proof of that was based on molecular outflow studies where the outflow rates consistently increase with increasing clump mass and luminosity of the region (Sect.~\ref{outflows}). Since the outflow rates are directly related to the accretion rates, one infers higher accretion rates for forming more massive stars. In a similar direction, studies of infall on clump and core scales also indicate that the infall rates are larger for high-mass regions compared to their low-mass counterparts.

Further characteristic parameters are the volume and column densities required to form either low- or high-mass stars. As outlined in section \ref{thresholds}, there are clear quantitative differences between the volume and column densities where active star formation is found in the two regimes. These thresholds vary by almost an order of magnitude. This implies that not just the mass reservoir available is responsible for star formation (also the Taurus cloud has a large mass of $\sim$2.4$\times 10^4$\,M$_{\odot}$, \citealt{goldsmith2008}), but that the gas has to be at higher volume and column densities for high-mass star formation. Hence, compression of the gas to higher densities is needed to make a cloud capable to form high-mass stars. In a similar direction, the mean separations between cores in forming clusters is smaller in high-mass regions compared to their low-mass counterparts (section \ref{fragmentation}). The most straightforward interpretation of that is again the higher densities that result in smaller thermal Jeans-lengths and by that smaller separations for the high-mass regime.

Regarding chemical effects, observations point to a similar prestellar composition of the gas and dust in low- and high-mass star-forming regions. The higher temperatures and energy output of high-mass protostars induce a stronger and spatially more extended impact on the chemical properties and detectability of complex molecules in the environment of the star-forming regions. 

The probably still least explored aspect is the contribution and importance of the magnetic field to the star formation process in the different mass regimes. As outlined in section \ref{magnetic}, there is tentative evidence that on large cloud formation scales the magnetic fields should be higher in high-mass star-forming regions because without magnetic support it may be difficult to form the observed massive filamentary structures at all. When moving to smaller spatial scales, the larger gravitational potential of the dense high-mass star-forming regions then overcomes more strongly the magnetic support resulting in higher mass-to-magnetic-flux ratios $(M/\Phi_B)$ for the more massive regions. In that picture, the importance of the magnetic field for low- and high-mass star formation may also be scale-dependent.

\subsection{Qualitative differences}

Clear qualitative difference between low- and high-mass star formation can be found in a few characteristics: strong and more energetic radiation from high-mass protostars and the associated feedback processes (Sect.~\ref{feedback}), as well as the much larger fraction of multiple systems in the high-mass regime (Sects.~\ref{environment} and \ref{fragmentation}).

Regarding the more energetic radiation, it can affect the immediate environment of the protostar by the formation of hypercompact H{\sc ii} regions, but it can also affect the slightly larger-scale environment when expanding H{\sc ii} regions form. These H{\sc ii} regions may either destroy the parental gas clump (destructive feedback), but they may also trigger new star formation events along the edges (constructive feedback). Although important by themselves, the large-scale constructive and destructive processes are unlikely to affect the actual processes forming low- and high-mass stars in these environments differently. More important in the context of this review are hence the small-scale hypercompact H{\sc ii} regions (Sect.~\ref{feedback}). The early assumption that the accretion should stop as soon as such hyper- and ultracompact H{\sc ii} regions form was soon challenged by observations of trapped H{\sc ii} regions and ongoing ionized accretion in rotating disk-like structures. Therefore, it appears that although the ionizing radiation is indeed qualitatively different between the low- and high-mass regime (almost absent in low-mass regions), the actual accretion processes in rotating disk-like structures may even stay similar with ionizing radiation. Hence, while low-mass as well as young high-mass disks are in general largely molecular, more evolved high-mass regions can continue disk accretion, just via an ionized accretion flow.

The bloating of protostars (Sect.~\ref{feedback}) may also become important in that context. Bloating protostars have correspondingly lower surface temperatures and hence significantly reduced UV radiation. Hence, the bloating of protostars, caused by the high accretion rates in high-mass star-forming regions, may significantly delay the formation of hypercompact H{\sc ii} regions and by that allow the star formation processes to continue longer to much higher masses without the complication of the ionizing radiation.

As outlined in Sect.~\ref{fragmentation}, the higher multiplicity around high-mass stars is a clear difference for the two regimes. The currently most likely explanation for at least the small-separation multiples is that the larger and more massive disks around high-mass protostars are more prone to fragmentation and hence lead to more multiple systems. Since the change in multiplicity is not a step-function but rather follows a slope, one may also consider that a quantitative and not qualitative change (Sect.~\ref{quantitative}). 

\begin{figure}[ht]
\includegraphics[width=0.99\textwidth]{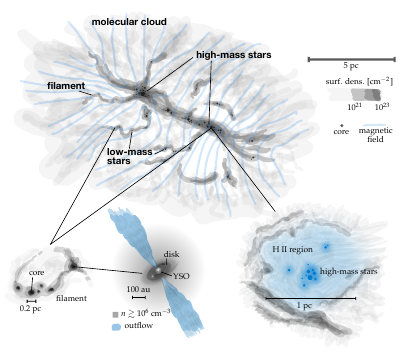}
\caption{Sketch of a star formation complex encompassing low- and high-mass star formation. While the top-part presents a large, magnetized filamentary cloud with low- and high-mass star formation occurring in different density regimes, the insets in the bottom part outline sub-aspects. While cores, disks and jets can be found in low- and high-mass regions, the feedback processes exclusively stem from high-mass stars. The figure is inspired by observational and numerical data from \citet{grudic2022,soler2019,traficante2023,oliva2023b}, and a Hubble image of Orion (credit: NASA, C.R.~O'Dell and S.K.~Wong). The entire figure is created by André Oliva.}
\label{sketch_summary}
\end{figure}

\subsection{Toward a unified description of low- and high-mass star formation}

Since its beginnings, the
study of star formation has had to face the enormous complexity of the 
physics that governs the transition from diffuse 
interstellar material into long-lived nuclear-fusing stars.
To tackle this complexity, it has been necessary to break up the problem into simpler pieces that could be individually managed and more easily solved. 
While this approach has been extremely successful, it has also led to a
fragmented community of researchers that study the formation of either low- or high-mass stars, and that not always have kept a fluid communication.  
As we have seen in the previous sections, low- and high- mass stars do show notable differences in their formation, but these differences are more often 
a matter of degree than of qualitative nature, especially concerning the earliest phases of the process.  Indeed, low- and high-mass stars often form together in large groups as a result of the fragmentation of a single cloud or clump of gas, indicating that the initial conditions of star formation can be the same for stars of very different masses.  In addition, the resulting stars follow a smoothly-varying, quasi-universal IMF, suggesting that their formation process varies continuously with mass.
%The resulting stars follow a smoothly-varying, quasi-universal IMF that also points to a full continuity between the product of the formation of low- and high-mass stars.

The realization that low- and high-mass stars often form in conjunction and follow similar developmental paths points to the need for a more 
unified approach in the study of star formation. Instruments like ALMA, NOEMA or JWST are sensitive enough to detect for the first time solar-mass stars at kpc distances, and therefore reveal how both low- and high-mass objects form simultaneously in embedded protoclusters.  In addition, extragalactic observations routinely provide a global picture of star formation where low- and high-mass stars are not individually distinguished (only high-mass stars are well detectable at extragalactic distances), but 
where the connection between their formation rate and the available molecular gas is strikingly tight (e.g., \citealt{kennicutt2012,schinnerer2024}).
If we are to properly interpret these and other observations, 
a unified description of star formation is more needed than ever.

While reaching a unified description of star formation is still a distant goal,
we hope that this review provides a motivation
to advance such an endeavor. Fig.~\ref{sketch_summary} presents a schematic 
picture of some of the elements that such a more unified description 
of star formation needs to 
combine, and which encompass dimensions of both space and time. 
In terms of space, a
unified description should connect the large-scale properties of a 
turbulent and magnetized cloud with the cascade of fragmentation responsible for 
the observed complex web of filaments, clumps, and cores. This web of structures sets 
the initial conditions for the formation of stars in different environments that range 
from isolated groups of low-mass stars to dense protoclusters 
with thousands of stars of different masses.
Once protostars have begun to form, disks and outflows around them naturally 
develop from the interplay between gravitational collapse, angular momentum and the magnetic field.
In regions of high stellar density, interactions between the newly-born stars may
occur and disturb the distribution of companions with which the stars were born. 
At a later stage, the combined action of outflows from stars of different masses 
plus the expansion of H{\sc ii} regions ionized by the high-mass stars will disrupt the 
cloud and eventually shut down the star-formation process. This
feedback, however, will combine with additional input from the 
resulting supernovae to trigger large-scale gas motions that will 
gather material and form a new generation of molecular clouds, maintaining in this way a
continuous cycle of formation and destruction that has been active
across the interstellar medium for billions of years.

\section{Conclusions, Summary and Outlook}

We have converged on a picture from low- to high-mass star formation where the physical and chemical processes over the whole mass regime share a lot of similarities but also exhibit some significant differences. We stress that star formation typically happens in a clustered mode, independent of mass. Figure \ref{sketch_summary} presents a sketch how low- and high-mass star formation may proceed within the same large-scale cloud environments, but where the high-mass processes occur in preferred high-density regions. Processes like core fragmentation, disk or outflow formation happen during the formation of stars of all masses, whereas radiation feedback is limited to the high-mass regions. The main conclusions of our review can be summarized as:

\begin{summary}[SUMMARY POINTS]
\begin{enumerate}
\item Some characteristics like the turbulent gas properties on clump/core scales or the density structures of the star-forming regions do not exhibit strong differences from low to high masses. Hence, turbulent support does not appear as a significant discriminant between low- and high-mass star formation. Furthermore, parameters like the IMF or stellar multiplicity appear to not be critically dependent on the environment.

\item While the absolute timescales of the main accretion phases are similar for protostars of all masses, the related accretion and outflow rates increase significantly by orders of magnitude from low- to high-mass regions. In addition to that, mean column and volume densities also increase for the high-mass regime. The latter directly results in smaller core separations for more massive regions. The higher densities and temperatures around high-mass protostars typically induce stronger emission from complex molecules. Furthermore, multiplicity strongly increases from low- to high-mass stars. While the small-scale higher multiplicity for high-mass protostars may indeed need additional disk fragmentation processes, the larger separation multiplicities appear to be explainable by core fragmentation on cluster scales. So far less understood aspects relate to the importance of the magnetic field. While on small scales gravity overcomes the magnetic support, magnetic fields may be more important in the high-mass regime during the formation of the large filamentary structures. 

\item The most obvious differences relate to the ionizing radiation exclusively stemming from the high-mass protostars. Although this ionizing radiation clearly impacts the environment by constructive and destructive feedback, studies indicate that accretion can still continue via ionized accretion flows. 

\end{enumerate}
\end{summary}

The analyses of the presented observations and theory reveal many open questions that need to be addressed by future studies in order to reach a unified view of star formation. To highlight a few:

\begin{issues}[FUTURE ISSUES]
\begin{enumerate}
\item How important is the magnetic field? Is a strong magnetic field on cloud scales required to form stable large and massive filamentary structures? Furthermore, the influence of the magnetic field on the fragmentation properties of the star-forming clumps needs deeper investigations.
\item A quantitative understanding of the accretion flow from large cloud-scales via clumps and cores onto individual protostars is needed. On the smallest spatial scales that relates to detailed studies of the disk-outflow connection. 
%In this context, one also has to understand  how the accretion rates vary between low- and high-mass protostars within the same clusters. 
\item Quantifying accretion variability and its origin(s) in detail is important over the entire mass range.
\item Detailed work on the multiplicity at birth
%compared to that of more evolved clusters 
is important to better understand the multiplicity differences between low- and high-mass regions. The role of disk fragmentation has to be investigated in greater depth.
\item How is turbulence dissipated on different scales? Does the impact along the cascade of scales change from low- to high-density regions?
\item How do processes in low-mass star formation vary between isolated Taurus-like regions compared to massive clusters with intense feedback processes from high-mass protostars?
\item Feedback in forming clusters needs to be understood better. How important are the different feedback processes to terminate the star formation activity?
%, e.g.,  by %exhausting the gas content of the region or stopping the global collapse? 
And how do feedback processes affect disk and planet formation?
\item How different is the formation of low- and high-mass stars in the typical Galactic regions discussed in this review compared to that in extreme environments like the CMZ and galactic mergers? Can the differences in environment explain the top-heavy IMFs reported in some observations?
\end{enumerate}
\end{issues}

\section*{DISCLOSURE STATEMENT}
The authors are not aware of any affiliations, memberships, funding, or financial holdings that might be perceived as affecting the objectivity of this review. 

\section*{ACKNOWLEDGMENTS}
We like to thank several people for inspiring discussions we had in the process of writing that review. 
In particular, we like to thank 
Vardan Elbakyan, 
Caroline Gieser, 
Thomas Haworth, 
Doug Johnstone, 
Hendrik Linz, 
Mordecai-Mark Mac Low, 
Tom Megeath, 
Selma de Mink, 
Phil Myers, 
Stella Offner, 
André Oliva, 
Thushara Pillai,
Ralph Pudritz, and
Qizhou Zhang.
We also thank Ewine van Dishoeck and Neal Evans for detailed comments on the draft.
Fig.~\ref{p_index} is provided by Caroline Gieser. Fig.~\ref{sketch_summary} is created by André Oliva.
RK acknowledges financial support via the Heisenberg Research Grant funded by the Deutsche Forschungsgemeinschaft (DFG, German Research Foundation) under grant no.~KU 2849/9, project no.~445783058. 

\def\aj{AJ}%
          % Astronomical Journal
\def\araa{ARA\&A}%
          % Annual Review of Astron and Astrophys
\def\apj{ApJ}%
          % Astrophysical Journal
\def\apjl{ApJ}%
          % Astrophysical Journal, Letters
\def\apjs{ApJS}%
          % Astrophysical Journal, Supplement
\def\ao{Appl.~Opt.}%
          % Applied Optics
\def\apss{Ap\&SS}%
          % Astrophysics and Space Science
\def\aap{A\&A}%
          % Astronomy and Astrophysics
\def\aapr{A\&A~Rev.}%
          % Astronomy and Astrophysics Reviews
\def\aaps{A\&AS}%
          % Astronomy and Astrophysics, Supplement
\def\azh{AZh}%
          % Astronomicheskii Zhurnal
\def\baas{BAAS}%
          % Bulletin of the AAS
\def\jrasc{JRASC}%
          % Journal of the RAS of Canada
\def\memras{MmRAS}%
          % Memoirs of the RAS
\def\mnras{MNRAS}%
          % Monthly Notices of the RAS
\def\pra{Phys.~Rev.~A}%
          % Physical Review A: General Physics
\def\prb{Phys.~Rev.~B}%
          % Physical Review B: Solid State
\def\prc{Phys.~Rev.~C}%
          % Physical Review C
\def\prd{Phys.~Rev.~D}%
          % Physical Review D
\def\pre{Phys.~Rev.~E}%
          % Physical Review E
\def\prl{Phys.~Rev.~Lett.}%
          % Physical Review Letters
\def\pasp{PASP}%
          % Publications of the ASP
\def\pasj{PASJ}%
          % Publications of the ASJ
\def\qjras{QJRAS}%
          % Quarterly Journal of the RAS
\def\skytel{S\&T}%
          % Sky and Telescope
\def\solphys{Sol.~Phys.}%
          % Solar Physics
\def\sovast{Soviet~Ast.}%
          % Soviet Astronomy
\def\ssr{Space~Sci.~Rev.}%
          % Space Science Reviews
\def\zap{ZAp}%
          % Zeitschrift fuer Astrophysik
\def\nat{Nature}%
          % Nature
\def\iaucirc{IAU~Circ.}%
          % IAU Cirulars
\def\aplett{Astrophys.~Lett.}%
          % Astrophysics Letters
\def\apspr{Astrophys.~Space~Phys.~Res.}%
          % Astrophysics Space Physics Research
\def\bain{Bull.~Astron.~Inst.~Netherlands}%
          % Bulletin Astronomical Institute of the Netherlands
\def\fcp{Fund.~Cosmic~Phys.}%
          % Fundamental Cosmic Physics
\def\gca{Geochim.~Cosmochim.~Acta}%
          % Geochimica Cosmochimica Acta
\def\grl{Geophys.~Res.~Lett.}%
          % Geophysics Research Letters
\def\jcp{J.~Chem.~Phys.}%
          % Journal of Chemical Physics
\def\jgr{J.~Geophys.~Res.}%
          % Journal of Geophysics Research
\def\jqsrt{J.~Quant.~Spec.~Radiat.~Transf.}%
          % Journal of Quantitiative Spectroscopy and Radiative Trasfer
\def\memsai{Mem.~Soc.~Astron.~Italiana}%
          % Mem. Societa Astronomica Italiana
\def\nphysa{Nucl.~Phys.~A}%
          % Nuclear Physics A
\def\physrep{Phys.~Rep.}%
          % Physics Reports
\def\physscr{Phys.~Scr}%
          % Physica Scripta
\def\planss{Planet.~Space~Sci.}%
          % Planetary Space Science
\def\procspie{Proc.~SPIE}%
          % Proceedings of the SPIE
\let\astap=\aap
\let\apjlett=\apjl
\let\apjsupp=\apjs
\let\applopt=\ao

\bibliography{bibliography}   
\bibliographystyle{ar-style2}    % this does the style, ar-style2.bst necessary

\end{document}